\documentclass[journal]{IEEEtran}

\usepackage{amsthm}
\usepackage{amsmath}
\usepackage{amsfonts}
\usepackage{amssymb}

\usepackage{bm}
\usepackage{graphicx}
\usepackage{subfigure}
\usepackage{stfloats}
\usepackage{booktabs}
\usepackage{hyperref}
\usepackage{cite}
\usepackage{tabularx}
\usepackage{textcomp}
\usepackage[ruled,linesnumbered,vlined]{algorithm2e}

\begin{document}

\title{Robust Secrecy via Aerial Reflection and Jamming: Joint Optimization of Deployment and Transmission}

\author{Xiao~Tang,~
        Hongliang~He,~
        Limeng Dong,~
        Lixin~Li,~
        Qinghe Du,~
        and~Zhu~Han
\thanks{X. Tang, L. Dong, and L. Li are with the School of Electronics and Information, Northwestern Polytechnical University, Xi'an 710072, China. (Email: tangxiao@nwpu.edu.cn)}
\thanks{H. He is with the School of Mechanical Engineering and Electronic Information, China University of Geosciences, Wuhan 430074, China.}
\thanks{Q. Du is with the Department of Communication Engineering, Xi'an Jiaotong University, Xi'an 710049, China.}
\thanks{Z. Han is with the Department of Electrical and Computer Engineering, University of Houston, Houston TX, 77004, USA.}
}

\maketitle

\begin{abstract}
Reconfigurable intelligent surfaces (RISs) are recognized with great potential to strengthen wireless security, yet the performance gain largely depends on the deployment location of RISs in the network topology. In this paper, we consider the anti-eavesdropping communication established through a RIS at a fixed location, as well as an aerial platform mounting another RIS and a friendly jammer to further improve the secrecy. The aerial RIS helps enhance the legitimate signal and the aerial cooperative jamming is strengthened through the fixed RIS. The security gain with aerial reflection and jamming is further improved with the optimized deployment of the aerial platform. We particularly consider the imperfect channel state information issue and address the worst-case secrecy for robust performance. The formulated robust secrecy rate maximization problem is decomposed into two layers, where the inner layer solves for reflection and jamming with robust optimization, and the outer layer tackles the aerial deployment through deep reinforcement learning. Simulation results show the deployment under different network topologies and demonstrate the performance superiority of our proposal in terms of the worst-case security provisioning as compared with the baselines.
\end{abstract}


\IEEEpeerreviewmaketitle

\section{Introduction} \label{sec1}

Reconfigurable intelligent surface (RIS) is envisioned as a paradigm-shifting technology to empower the next-generation wireless communications. With a massive number of low-cost and low-power reflecting elements to alter the electromagnetic properties of the incident signal, RIS enables controllable and programmable wireless propagation rather than only adapting to the environments as conventional communications~\cite{ris}. In this regard, RIS-assisted communication features to intentionally add up the received signals constructively or destructively to improve the desired receptions while weakening the unintended ones. Due to the cost-effective and flexible operations of RISs, there has emerged rich literature investigating various aspects of RIS-enabled communications, e.g., RIS channel modeling, RIS-assisted transmissions, RIS-enhanced information security, etc.~\cite{risCom}.

Security is one of the primary concerns for wireless communications, for which physical layer security featuring keyless operations while providing information-theoretical secrecy has arisen as an attracting solution~\cite{sec}. Physical layer security technique defends against malicious adversaries by exploiting the randomness of wireless medium, in this respect, the ability of RISs to actively intervene the wireless environment has provided an additional degree of freedom to further enhance the information security~\cite{risSec}. With RIS-enabled intelligent radio, we can intentionally improve the legitimate reception while reducing the signal leakages to unintended third parties and thus improves the security performance significantly. Therefore, RIS-enhanced security has attracted wide attention recently, including RIS-assisted friendly relaying, jamming mitigation, artificial noise design, etc.~\cite{risEve,risSecApp}.

Despite the potential of RISs to enhance wireless communications, the performance gain is heavily affected by the RIS-channel quality. In particular, the RIS-related channel fundamentally depends on the product of the incident signal channel, reflection channel, and the phase shifts, and thus the deployment of RISs has a great impact on the overall performance~\cite{risDeploy}. For example, although it is more likely to establish line-of-sight links via RISs deployed on high-rise buildings, the overall transmission distance over RISs is usually rather longer as compared with the direct links, and thus may not be able to provide desired performance enhancement. Towards this issue, a direct complement is to increase the number of reflecting elements or deploy more RISs, yet this can be severely restricted by the physical conditions~\cite{doubleRis,Li1}. Also, one may resort to new-type RISs with more desired properties, such as active RISs with amplified reflection, whereas it essentially relies on the advance of material or circuit design that largely goes beyond the scope of conventional communications~\cite{activeRis}.

Recently, rapid progress has been witnessed in various aerial platforms, such as unmanned aerial vehicles (UAVs), high altitude platforms (HAPs), and airships~\cite{Xiong1,Li2}. We can then exploit the aerial platforms as the base for RISs, leading to the aerial RISs (ARISs) that enable flexibly deployed RISs to achieve the optimized reflection in various network topology~\cite{aRis,risAccess}. Regarding RIS-assisted physical layer security, ARISs can significantly strengthen the legitimate signal and downgrade the eavesdropping. Particularly, through reflection at the optimal location, we can effectively weaken the signal leakage or enhance the artificial jamming at unintended receivers thus improving the secrecy rate~\cite{arisSec,arisPhySec}. Therefore, the ARISs with flexible and on-demand deployment have great advantages as compared with fixed RISs, and have tremendous potential to catalyze conventional security approaches toward more adequately-protected information security.

Attracted by the RIS-benefited wireless security, we in this paper propose to deploy a fixed RIS as well as an ARIS for anti-eavesdropping communications. The aerial platform carrying the ARIS is also associated with a friendly jammer, enhancing the secrecy through cooperative reflection and jamming. We particularly consider the cases that the channels related to the eavesdroppers are associated with uncertainties. By exploiting robust optimization and learning techniques, we propose a joint design of aerial deployment, reflection at the RISs, and jamming, to maximize the robust secrecy. To summarize, the main contributions are highlighted as follows:
\begin{itemize}
	\item We propose to deploy an aerial platform carrying an ARIS and a cooperative jammer, along with a fixed RIS, to enhance wireless secrecy. The aerial reflection and jamming are enabled with flexible deployment so as to coordinate with the fixed RIS, improving the legitimate transmissions while downgrading the eavesdropping.
	\item We employ the cascaded reflection channel model and consider the imperfect channel state information at the eavesdroppers. Targeting at the worst case for robustness, we formulate the problem to maximize the robust secrecy by jointly considering the aerial deployment, reflection at the RISs, and jamming optimization.
	\item We decompose the problem into two layers, where the inner layer optimizes the secure transmission and the outer layer for deployment. In the inner layer, by deriving the worst-case secrecy rate, the reflection and jamming strategies are obtained within a block coordinate descent (BCD) framework. Then, the outer-layer deployment is obtained with deep reinforcement learning technique.
\end{itemize}

The rest of this paper is organized as follows. In Sec.~\ref{sec2}, we review the related work. In Sec.~\ref{sec3}, we present the system model with the robust secrecy optimization problem formulation. In Sec.~\ref{sec4}, the inner transmission problem is solved to optimize reflection and jamming. In Sec.~\ref{sec5}, the deployment of the aerial platform is tackled with deep reinforcement learning. Sec.~\ref{sec6} provides the simulation results, and finally Sec.~\ref{sec7} concludes this paper.

\section{Related Work} \label{sec2}

Due to the ability to actively intervene the signal propagation, RISs have the potential to enhance wireless communications in various aspects~\cite{ris,risCom}. Particularly, the interplay between RISs and physical layer security has shown significant advantages to safeguarding secrecy while defending against eavesdropping~\cite{risSec}. In~\cite{risSecApp}, the authors consider the artificial noise-aided secure communications, where a RIS is invoked to enhance the secrecy rate by jointly optimizing the reflection and jamming. In~\cite{risNoma}, the authors consider the non-orthogonal multiple access scenario with a RIS to assist the secure transmissions. In~\cite{risMec}, the authors investigate the secure edge computing issue, where the reflection-enhanced transmission is jointly optimized with the computing strategy to secure the computation offloading. In~\cite{risOr}, the authors exploit a RIS as a backscatter device that modulates the received confidential signal to jamming signal to deteriorate the eavesdropping. In~\cite{risBoth}, the authors consider both eavesdropping and jamming attacks, where a RIS with reflection optimization is exploited to enhance security. In~\cite{risKey}, the authors propose a RIS-assisted key generation scheme by intervening in the propagation in harsh environments. In~\cite{risLearn}, the authors exploit the non-cooperative game to model the interaction between a RIS-assisted legitimate user and a smart attacker with learning-based security solutions.

Recently, the deployment of RISs has raised increasing interest with joint consideration of reflection-based transmissions. In~\cite{risAccess}, the authors jointly consider the RIS deployment and access strategies for maximized system rate. In~\cite{risFD}, the authors investigate the full-duplex system while jointly optimizing the passive beamforming and deployment of the RIS. In this aspect, the various aerial platforms, particularly UAVs, have enabled ARISs as a more flexible solution~\cite{arisSec}. In~\cite{aris3D}, the authors exploit an ARIS to maximize the worst-case signal-to-noise ratio by jointly considering the transmission and AIRS placement. In~\cite{arisMamimo}, the authors propose to deploy multiple ARISs forming a massive multiple-input multiple-output network to extend the network coverage. Moreover, ARIS-assisted security has also emerged as an attractive solution to defend against various attacks. In~\cite{arisPhySec}, the authors investigate various use cases to integrate UAVs and RISs to enhance physical layer security. In~\cite{arisICN}, the authors propose to use a UAV-carried RIS to defend against eavesdropping with trajectory optimization. In~\cite{arisJam}, the authors address the anti-jamming communications by leveraging the ARIS reflection and deployment. In above work, the RISs are either fixed or deployed with UAVs, where the former potentially lacks flexibility while the latter is of high cost. As such, we may resort to the on-demand and adaptive use of both types to tackle the unfavorable transmission scenario with reasonable expenditure.

As the reflection-based transmission through RISs raises higher challenges for channel measurement, many research efforts have been devoted to the imperfect channel information issue. In~\cite{roCas}, the authors adopt the cascaded reflection channel model with imperfection and minimize the energy consumption. In~\cite{roCui}, the authors propose a robust design regarding the instantaneous beamforming and quasi-static phase shifts with channel imperfection, adapting to the rapid channel variation. In~\cite{roDis}, the authors consider the randomly distributed channel errors and investigate the wireless information and power transfer issue under probabilistically robust constraints. In~\cite{roSecFd}, the authors address the secure full-duplex communications with channel uncertainties and optimize the worst-case achievable secrecy rate. In~\cite{roSecUAV}, the authors achieve robust secrecy for RIS-aided UAV communications, with a joint design of transmission and UAV trajectory. In~\cite{roSecDis}, the authors consider the secure multicast beamforming with RISs, the distributionally robustness is achieved against different distributions of channel errors. While these researches address different uncertainty models, the considered cases only incorporate one fixed RIS. In this regard, it is worth investigating the uncertainty issue in some more complicated scenarios incorporating multiple RISs with different types, providing insights for the more generalized use cases of RIS-assisted wireless security.

\section{System Model} \label{sec3}

\begin{figure}[t]
  \centering
  \includegraphics[width=7.0cm]{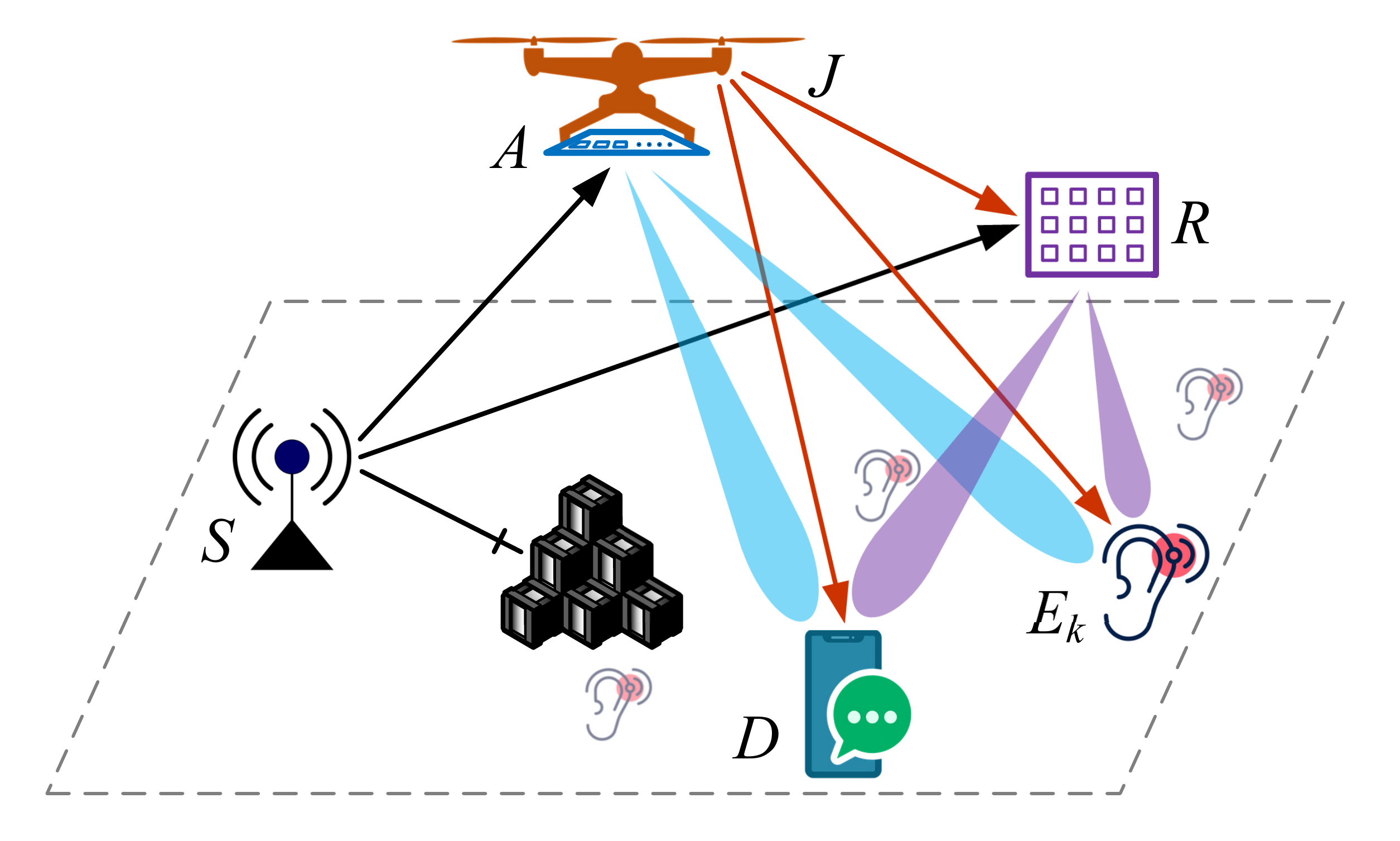}
  \caption{System model.}
  \label{fig:sys}
\end{figure}

We consider an area, denoted by $ \mathcal{A} $, with a legitimate source node, denoted by $ S $, having confidential information towards a legitimate destination node, denoted by $ D $. While in the same area, there are a set of eavesdroppers, denoted by $ E_k $ with $ k\in\mathcal{K}=\left\{1,2,\cdots, K\right\} $, intending to wiretap the legitimate transmissions, as shown in Fig.~\ref{fig:sys}. The legitimate source and destination, as well as the eavesdroppers, are assumed of one single antenna and are located on the ground. We consider the scenario that the legitimate source is blocked by certain high-rise obstacles, and thus there is no direct link towards the legitimate destination or eavesdroppers. Meanwhile, a RIS is deployed at a certain fixed location, noted as fixed RIS and denoted by $ R $, to establish reflection links to assist the transmission, where the reflected signals can also be overheard by the eavesdroppers. Suppose there are $ N_R $ reflection elements at $ R $, denoted by $ \mathcal{N}_R = \left\{1,2,\cdots,N_R\right\} $, the phase shifts are given as $ \bm{\theta}_R = \left[ \theta_{R,n}\right]_{n\in\mathcal{N}_R} $. Then the reflection-coefficient matrix is given as $ \bm{\Theta}_R = \mathsf{diag}\left( \bm{\vartheta}_R \right) $ with $ \bm{\vartheta}_R = \left[ \vartheta_{R,n}\right]_{n\in\mathcal{N}_R} $ and $ \vartheta_{R,n} = e^{j\theta_{R,n}} $. Also, the channels from $ S $ to $ R $, from $ R $ to $ D $, and from $ R $ to $ E_k $ are denoted by $ \bm{h}_{SR} $, $ \bm{h}_{RD} $, and $ \bm{h}_{Rk} $, respectively. Here, we assume the constant amplitude response to facilitate the analysis. A more practical and general model is proposed recently with phase-dependent amplitude~\cite{amp}. Our considered scenario can also incorporate such a reflection model, and as can be safely expected, our proposed framework can be extended to such cases with proper treatment of the amplitude issue.

Despite the transmissions established through the fixed RIS, it may still be challenging to guarantee the secrecy of legitimate communications. To this issue, we propose to deploy an aerial platform mounting a RIS and a jammer to enhance wireless secrecy. A typical scenario is that the fixed RIS is deployed at some high-rise buildings while not always as effective given the mobility or potentially unfavorable locations of the legitimate nodes. In this regard, an ARIS carried by a UAV is dispatched to assist the secure communications. The ARIS, denoted by $ A $, has geographic coordinates of $ \bm{w}_A = \left[ w_A^{\text{(x)}},w_A^{\text{(y)}}, H_A \right] $ with $ H_A $ being the fixed altitude, and enables communications through reflection in the air. Suppose there are $ N_A $ reflection elements at $ A $, denoted by $ \mathcal{N}_A = \left\{1,2,\cdots,N_A\right\} $ with corresponding phase shifts given as $ \bm{\theta}_A = \left[ \theta_{A,n}\right]_{n\in\mathcal{N}_A} $. Then the reflection-coefficient matrix is given as $ \bm{\Theta}_A = \mathsf{diag}\left( \bm{\vartheta}_A \right) $ with $ \bm{\vartheta}_A = \left[ \vartheta_{A,n}\right]_{n\in\mathcal{N}_A} $ and $ \vartheta_{A,n} = e^{j\theta_{A,n}} $. Similarly as the fixed RIS, the channels from $ S $ to $ A $, from $ A $ to $ D $, and from $ A $ to $ E_k $ are denoted by $ \bm{h}_{SA} $, $ \bm{h}_{AD} $, and $ \bm{h}_{Ak} $, respectively. We in this paper only consider the single-time reflection at the RISs, while the signals reflected multiple times or between the RISs are ignored due to the more severe attenuation therein. Moreover, a cooperative jammer, denoted by $ J $ with $ M $ antennas is also mounted onto the aerial platform as a flying helper emitting artificial noise to intentionally deteriorate the eavesdropping. Since the jammer belongs to the legitimate system, we assume that the legitimate signal and jamming signals are well coordinated. As the jammer is in the air, it has direct links to the legitimate destination and eavesdroppers. Then, the channel from $ J $ to $ D $ and from $ J $ to $ E_k $ are denoted by $ \bm{h}_{JD} $ and $ \bm{h}_{Jk} $, respectively. Also, the jamming signal reaches the ground through reflection. Given the physical space limitation in the aerial platform to carry ARIS and jamming device, e.g., jammer on top of the platform while ARIS beneath, we consider that the jamming signal is only reflected through the fixed RIS, and is not affected by the ARIS. The channel from $ J $ to $ R $ is denoted by $ \bm{H}_{JR} $, and the reflection link from $ R $ to the ground nodes are given as $ \bm{h}_{JD} $ and $ \bm{h}_{Jk} $.

For the considered system, the signals from the legitimate transmitter and jammer are denoted by $ x $ and $ \bm{z} $, respectively, where $ x\sim\mathcal{CN}\left(0, P_S\right) $ with $ P_S $ being the transmit power of $ S $, $ \bm{z}\sim\mathcal{CN}\left(\bm{0}_{M\times 1}, \bm{Z}\right) $ with $ \bm{Z} $ being the covariance of the jamming signal subject to the maximum jamming power specified by $ P_J $. Then the received signals at the legitimate destination $ D $, and eavesdropper $ E_k $, are
\begin{equation}
\begin{aligned}
        y_D =& \left( \bm{h}_{AD}^{\dag} \bm{\Theta}_A \bm{h}_{SA} + \bm{h}_{RD}^{\dag} \bm{\Theta}_R \bm{h}_{SR} \right) x \\
        &+ \left( \bm{h}_{JD}^{\dag} + \bm{h}_{RD}^{\dag} \bm{\Theta}_R \bm{H}_{JR} \right) \bm{z} + n_D,
\end{aligned}
\end{equation}
and
\begin{equation}
\begin{aligned}
        y_k =& \left( \bm{h}_{Ak}^{\dag} \bm{\Theta}_A \bm{h}_{SA} + \bm{h}_{Rk}^{\dag} \bm{\Theta}_R \bm{h}_{SR} \right) x \\
        &+ \left( \bm{h}_{Jk}^{\dag} + \bm{h}_{Rk}^{\dag} \bm{\Theta}_R \bm{H}_{JR} \right) \bm{z} + n_k,
\end{aligned}
\end{equation}
respectively, where $ n_D $ and $ n_k $ are the Gaussian noise at $ D $ and $ E_k $, respectively. In this paper, we adopt the cascaded channel model and rewrite the links established through reflection as
\begin{subequations} \label{eq:sub1}
\begin{equation}
	\bm{h}_{AD}^{\dag} \bm{\Theta}_A \bm{h}_{SA} = \bm{\vartheta}_A^{\dag} \underbrace{\mathsf{diag}\left( \bm{h}_{AD}^{\dag} \right)\bm{h}_{SA}}_{\overset{\Delta}{=}\bm{h}_{SAD}} = \bm{\vartheta}_A^{\dag} \bm{h}_{SAD},
\end{equation}
\begin{equation}
	\bm{h}_{RD}^{\dag} \bm{\Theta}_R \bm{h}_{SR} = \bm{\vartheta}_R^{\dag} \underbrace{\mathsf{diag}\left( \bm{h}_{RD}^{\dag} \right)\bm{h}_{SR}}_{\overset{\Delta}{=}\bm{h}_{SRD}} = \bm{\vartheta}_R^{\dag} \bm{h}_{SRD},
\end{equation}
\begin{equation}
	\bm{h}_{RD}^{\dag} \bm{\Theta}_R \bm{H}_{JR} = \bm{\vartheta}_R^{\dag} \underbrace{\mathsf{diag}\left( \bm{h}_{RD}^{\dag} \right)\bm{H}_{JR}}_{\overset{\Delta}{=}\bm{h}_{JRD}} = \bm{\vartheta}_R^{\dag} \bm{h}_{JRD},
\end{equation}
\end{subequations}
at legitimate receiver $ D $. Similarly, at eavesdropper $ E_k $, we have
\begin{subequations}  \label{eq:sub2}
\begin{equation}
	\bm{h}_{Ak}^{\dag} \bm{\Theta}_A \bm{h}_{Sk} = \bm{\vartheta}_A^{\dag} \underbrace{\mathsf{diag}\left( \bm{h}_{Ak}^{\dag} \right)\bm{h}_{SA}}_{\overset{\Delta}{=}\bm{h}_{SAk}} = \bm{\vartheta}_A^{\dag} \bm{h}_{SAk},
\end{equation}
\begin{equation}
	\bm{h}_{Rk}^{\dag} \bm{\Theta}_R \bm{h}_{SR} = \bm{\vartheta}_R^{\dag} \underbrace{\mathsf{diag}\left( \bm{h}_{Rk}^{\dag} \right)\bm{h}_{SR}}_{\overset{\Delta}{=}\bm{h}_{SRk}} = \bm{\vartheta}_R^{\dag} \bm{h}_{SRk},
\end{equation}
\begin{equation}
	\bm{h}_{Rk}^{\dag} \bm{\Theta}_R \bm{H}_{JR} = \bm{\vartheta}_R^{\dag} \underbrace{\mathsf{diag}\left( \bm{h}_{Rk}^{\dag} \right)\bm{H}_{JR}}_{\overset{\Delta}{=}\bm{H}_{JRk}} = \bm{\vartheta}_R^{\dag} \bm{H}_{JRk}.
\end{equation}
\end{subequations}

Based on the transmission model, the signal-to-interference-plus-noise ratio (SINR) at legitimate receiver $ D $, and eavesdropper $ E_k $, are
\begin{equation}
	\gamma_D = \frac{P_S \left| \bm{\vartheta}_A^{\dag} \bm{h}_{SAD} + \bm{\vartheta}_R^{\dag} \bm{h}_{SRD} \right|^2 }{ \left( \bm{h}_{JD}^{\dag} + \bm{\vartheta}_R^{\dag} \bm{h}_{JRD} \right) \bm{Z} \left( \bm{h}_{JD}^{\dag} + \bm{\vartheta}_R^{\dag} \bm{h}_{JRD} \right)^{\dag} + \sigma_0^2 },
\end{equation}
and
\begin{equation}
	\gamma_k = \frac{P_S \left| \bm{\vartheta}_A^{\dag} \bm{h}_{SAk} + \bm{\vartheta}_R^{\dag} \bm{h}_{SRk} \right|^2 }{ \left( \bm{h}_{Jk}^{\dag} + \bm{\vartheta}_R^{\dag} \bm{H}_{JRk} \right) \bm{Z} \left( \bm{h}_{Jk}^{\dag} + \bm{\vartheta}_R^{\dag} \bm{H}_{JRk} \right)^{\dag} + \sigma_0^2 },
\end{equation}
respectively, where $ \sigma_0^2 $ is the background noise power assumed identical at all receivers. Then, the secrecy rate is obtained as
\begin{equation}
	R_S = \left[ \log\left(1 + \gamma_D\right) - \max_{k\in\mathcal{K}} \log\left(1 + \gamma_k\right) \right]^+,
\end{equation}
where $ (\:\cdot\:)^+ = \max\{\:\cdot\:, 0\} $, and this operator is omitted for the discussions afterwards since the transmission will be ceased for negative secrecy rate.

Moreover, we consider that perfect channel state information can be obtained between the legitimate transmission pair, while the channel state information regarding the eavesdroppers is associated with errors. This is due to the fact that it is difficult to obtain precise channel information at the passive eavesdroppers, especially when we consider multiple eavesdroppers with reflection links. Correspondingly, the channels related to $ E_k $, $ \forall k\in\mathcal{K} $ are modeled as
\begin{subequations}  \label{eq:sub3}
\begin{equation} \label{eq:sakUnc}
	\bm{h}_{SAk} = \hat{\bm{h}}_{SAk} + \Delta\bm{h}_{SAk}, \text{  with  } \left\| \Delta\bm{h}_{SAk} \right\|^2 \le \epsilon_{SAk},
\end{equation}
\begin{equation} \label{eq:srkUnc}
	\bm{h}_{SRk} = \hat{\bm{h}}_{SRk} + \Delta\bm{h}_{SRk}, \text{  with  } \left\| \Delta\bm{h}_{SRk} \right\|^2 \le \epsilon_{SRk},
\end{equation}
\begin{equation} \label{eq:jkUnc}
	\bm{h}_{Jk} = \hat{\bm{h}}_{Jk} + \Delta\bm{h}_{Jk}, \text{  with  } \left\| \Delta\bm{h}_{Jk} \right\|^2 \le \epsilon_{Jk},
\end{equation}
\begin{equation} \label{eq:jrkUnc}
	\bm{H}_{JRk} = \hat{\bm{H}}_{JRk} + \Delta\bm{H}_{JRk}, \text{  with  } \left\| \Delta\bm{H}_{JRk} \right\|_F^2 \le \epsilon_{JRk},
\end{equation}
\end{subequations}
where $ \hat{\bm{h}}_{SAk} $, $ \hat{\bm{h}}_{SRk} $, $ \hat{\bm{h}}_{Jk} $, and $ \hat{\bm{H}}_{JRk} $ are estimation of the cascaded channels, $ \Delta{\bm{h}}_{SAk} $, $ \Delta{\bm{h}}_{SRk} $, $ \Delta{\bm{h}}_{Jk} $, and $ \Delta{\bm{H}}_{JRk} $ are the errors with $ \epsilon_{SAk} $, $ \epsilon_{SRk} $, $ \epsilon_{Jk} $, and $ \epsilon_{JRk} $ being the error bounds.

Based on the discussions above, we intend to maximize the secrecy rate of legitimate transmissions by jointly optimizing the artificial noise, reflection, and UAV deployment, in the presence of channel uncertainties at the eavesdroppers. Correspondingly, the secrecy optimization problem is formulated as
\begin{IEEEeqnarray}{cl}
	\IEEEyesnumber\label{eq:problem} \IEEEyessubnumber*
	\max_{\left[w_A^{\text{(x)}},w_A^{\text{(y)}}\right],\bm{\vartheta}_A,\bm{\vartheta}_R, \bm{Z}} \:\: & \min_{\Delta\bm{h}_{SAk}, \Delta\bm{h}_{SRk}, \Delta\bm{h}_{Jk}, \Delta\bm{H}_{JRk}} R_S \\
	{\:\:\:\rm{s.t.}} \quad & \left[w_A^{\text{(x)}},w_A^{\text{(y)}}\right] \in\mathcal{A}, \label{eq:wConstr} \\
	 & \left| \vartheta_{A,n} \right| = 1 ,\quad \forall n\in\mathcal{N}_A, \label{eq:AConstr} \\
	 & \left| \vartheta_{R,n} \right| = 1 ,\quad \forall n\in\mathcal{N}_R, \label{eq:RConstr} \\
	 & \mathsf{Tr}\left(\bm{Z}\right) \ge P_J, \quad \bm{Z}\succcurlyeq 0. \label{eq:zConstr}
\end{IEEEeqnarray}
The formulated problem is rather complicated with three-fold difficulties. First, the reflection optimization needs to consider the legitimate signal and artificial noise simultaneously, whose reflections are in an asymmetry manner as the legitimate signal is reflected by both RISs while the jamming signal is only reflected by the fixed RIS. Second, the mobility of the aerial platform affects the aerial reflection as well as friendly jamming, and further interplays with the reflection at the fixed RIS. Third, the channels at eavesdroppers are associated with uncertainties, which influence the reflection and jamming and need to be tackled to achieve robust secrecy.

In order to solve the problem effectively, we propose to decompose it into two layers, the inner layer optimizes the friendly jamming and reflection, while the outer layer tackles the aerial deployment. The decomposition is based on the fact that the inner-layer problem solving is conducted based on given channel conditions (though with uncertainties), while the outer-layer deployment affects the channel and further impacts the transmissions. Specifically, the deployment affects the system topology and thus the changes in large-scale channel conditions dominate the influence on system performance. Meanwhile, when the deployment is given, indicating fixed large-scale fading in the system, the transmission issue mainly addresses the small-scale fading along with the information uncertainties. Further, the inner and outer subproblems are solved through robust optimization and learning techniques, respectively, as elaborated in the following sections.

\section{Jamming and Reflection Optimization} \label{sec4}

In this section, we consider the inner problem to optimize the jamming and reflection with fixed UAV deployment, in the presence of channel uncertainties, specified as
\begin{IEEEeqnarray}{cl}
	\IEEEyesnumber\label{eq:innerProb} \IEEEyessubnumber*
	\max_{\bm{\vartheta}_A,\bm{\vartheta}_R, \bm{Z}} \:\: & \min_{\Delta\bm{h}_{SAk}, \Delta\bm{h}_{SRk}, \Delta\bm{h}_{Jk}, \Delta\bm{H}_{JRk}} R_S \\
	 {\:\:\:\rm{s.t.}} \quad & \left| \vartheta_{A,n} \right| = 1 ,\quad \forall n\in\mathcal{N}_A, \label{eq:inAConstr} \\
	 & \left| \vartheta_{R,n} \right| = 1 ,\quad \forall n\in\mathcal{N}_R, \label{eq:inRConstr} \\
	 & \mathsf{Tr}\left(\bm{Z}\right) \ge P_J, \quad \bm{Z}\succcurlyeq 0. \label{eq:inzConstr}
\end{IEEEeqnarray}
In~(\ref{eq:innerProb}), we can see that the reflection and jamming are complicatedly coupled with each other while jointly affected by the uncertainties. In this regard, we first tackle the uncertainties within the minimization operation in the form of explicit constraints. Then, we investigate the cooperative jamming, ARIS reflection, and fixed-RIS reflection separately according to the physical functionalities. We adopt the block coordinate descent (BDC) framework to solve the subproblems independently with different optimization techniques as detailed below.

\subsection{Reformulation Against Uncertainties} \label{sec:IV-A}

For the considered problem of secrecy enhancement, the uncertainties are associated with the channels at the eavesdroppers and further affect the jamming and reflection strategies. Mathematically, the uncertainties are incorporated in the SINRs at the eavesdropper as part of the objective function in the formulated problem, which is tackled first to facilitate the analyses. In this respect, by introducing new variables $ \left\{ \psi_{Sk} \right\}_{k\in\mathcal{K}} $ and $ \left\{ \psi_{Jk} \right\}_{k\in\mathcal{K}} $ with
\begin{equation} \label{eq:psiS}
	\left| \bm{\vartheta}_A^{\dag} \bm{h}_{SAk} + \bm{\vartheta}_R^{\dag} \bm{h}_{SRk} \right|^2 \le \psi_{Sk}, \quad\forall k\in\mathcal{K},
\end{equation}
and
\begin{equation} \label{eq:psiJ}
	\left( \bm{h}_{Jk}^{\dag} + \bm{\vartheta}_R^{\dag} \bm{H}_{JRk} \right) \bm{Z} \left( \bm{h}_{Jk}^{\dag} + \bm{\vartheta}_R^{\dag} \bm{H}_{JRk} \right)^{\dag} \ge \psi_{Jk},\quad\forall k\in\mathcal{K},
\end{equation}
we reach the inequalities regarding SINRs at the eavesdroppers as
\begin{equation} \label{eq:gammaUp}
	\gamma_k \le \frac{P_S \psi_{Sk}}{\psi_{Jk} + \sigma_0^2}, \quad\forall k\in\mathcal{K}.
\end{equation}
Specifically, the inequality in~(\ref{eq:psiS}) can be reinterpreted through Schur complement as
\begin{equation} \label{eq:schurS}
	\begin{bmatrix}
	\psi_{Sk} & \left(\bm{\vartheta}_A^{\dag} \bm{h}_{SAk} + \bm{\vartheta}_R^{\dag} \bm{h}_{SRk}\right)^{\dag} \\
	\bm{\vartheta}_A^{\dag} \bm{h}_{SAk} + \bm{\vartheta}_R^{\dag} \bm{h}_{SRk} & 1
	\end{bmatrix}
	\succcurlyeq 0.
\end{equation}
Then, by substituting the channels with uncertainties in~(\ref{eq:sakUnc}) and~(\ref{eq:srkUnc}) into the Schur complement condition in~(\ref{eq:schurS}), we have the following inequality along with a further derivation as
\begin{equation} \label{eq:schurEx}
\begin{aligned}
&	\begin{bmatrix}
	\psi_{Sk} & \left(\bm{\vartheta}_A^{\dag} \hat{\bm{h}}_{SAk} + \bm{\vartheta}_R^{\dag} \hat{\bm{h}}_{SRk}\right)^{\dag} \\
	\bm{\vartheta}_A^{\dag} \hat{\bm{h}}_{SAk} + \bm{\vartheta}_R^{\dag} \hat{\bm{h}}_{SRk} & 1
	\end{bmatrix} \\
	\succcurlyeq
&	\begin{bmatrix}
	0 & \left(\bm{\vartheta}_A^{\dag} \Delta\bm{h}_{SAk} \right)^{\dag} \\
	\bm{\vartheta}_A^{\dag} \Delta\bm{h}_{SAk} & 0
	\end{bmatrix} \\
	&+
	\begin{bmatrix}
	0 & \left(\bm{\vartheta}_R^{\dag} \Delta\bm{h}_{SRk}\right)^{\dag} \\
	\bm{\vartheta}_R^{\dag} \Delta\bm{h}_{SRk} & 0
	\end{bmatrix} \\
	=
&	-\begin{bmatrix} 1 \\ 0 \end{bmatrix} \Delta\bm{h}_{SAk}^{\dag} \begin{bmatrix} \bm{0}_{N_A\times 1} \:\: \bm{\vartheta}_A \end{bmatrix} - 
\begin{bmatrix} \bm{0}_{1\times N_A} \\ \bm{\vartheta}_A^{\dag} \end{bmatrix}
\Delta\bm{h}_{SAk} \begin{bmatrix} 1 \:\: 0 \end{bmatrix} \\
&	-\begin{bmatrix} 1 \\ 0 \end{bmatrix} \Delta\bm{h}_{SRk}^{\dag} \begin{bmatrix} \bm{0}_{N_R\times 1} \:\: \bm{\vartheta}_R \end{bmatrix} - 
\begin{bmatrix} \bm{0}_{1\times N_R} \\ \bm{\vartheta}_R^{\dag} \end{bmatrix}
\Delta\bm{h}_{SRk} \begin{bmatrix} 1 \:\: 0 \end{bmatrix} 
\end{aligned},
\end{equation}
which separates the channel estimations and uncertainties on the left-hand side and right-hand side, respectively. The inequality in~(\ref{eq:schurEx}) facilitates the application of general sign-definiteness principle~\cite{signdef}, leading to the following equivalent inequality as~(\ref{eq:signDef}),
\begin{figure*}[t]
\begin{equation} \label{eq:signDef}
        \begin{bmatrix}
        \psi_{Sk} - \rho_{1,k} - \rho_{2,k} & \left(\bm{\vartheta}_A^{\dag} \hat{\bm{h}}_{SAk} + \bm{\vartheta}_R^{\dag} \hat{\bm{h}}_{SRk}\right)^{\dag} & \bm{0}_{1\times N_A} & \bm{0}_{1\times N_R} \\
        \bm{\vartheta}_A^{\dag} \hat{\bm{h}}_{SAk} + \bm{\vartheta}_R^{\dag} \hat{\bm{h}}_{SRk} & 1 & \sqrt{\epsilon_{SAk}}\bm{\vartheta}_A^{\dag} & \sqrt{\epsilon_{SRk}}\bm{\vartheta}_R^{\dag} \\
        \bm{0}_{N_A\times 1} & \sqrt{\epsilon_{SAk}}\bm{\vartheta}_A & \rho_{1,k}\bm{I}_{N_A \times N_A} & \bm{0}_{N_A\times N_R} \\
        \bm{0}_{N_R\times 1} & \sqrt{\epsilon_{SRk}}\bm{\vartheta}_R & \bm{0}_{N_R\times N_R} & \rho_{2,k}\bm{I}_{N_R \times N_R}
        \end{bmatrix}
        \succcurlyeq 0,
\end{equation}
\hrulefill
\end{figure*}
where $ \left\{\rho_{1,k}\right\}_{k\in\mathcal{K}} $ and $ \left\{\rho_{2,k}\right\}_{k\in\mathcal{K}} $ are the non-negative variables newly introduced along with the general sign-definiteness principle. In~(\ref{eq:signDef}), the uncertainty parts are replaced with the corresponding error bound, indicating that the inequality in~(\ref{eq:signDef}) acts as the robust counterpart for the inequality in~(\ref{eq:psiS}). Moreover, the inequality in~(\ref{eq:signDef}) incorporates reflection coefficients and introduced auxiliaries in the form of linear matrix inequalities, which are convex and can be conveniently tackled in existing solvers.

To deal with the constraints in~(\ref{eq:psiJ}), we first introduce the following reformulation
\begin{equation} \label{eq:psiJTilde}
	\bm{h}_{Jk}^{\dag} + \bm{\vartheta}_R^{\dag} \bm{H}_{JRk} = \underbrace{\left[ 1 \:\: \bm{\vartheta}_R^{\dag} \right]}_{\overset{\Delta}{=}\tilde{\bm{\vartheta}}_R^{\dag}} \underbrace{\left[ \begin{array}{c} \bm{h}_{Jk}^{\dag} \\ \bm{H}_{JRk} \end{array} \right]}_{\overset{\Delta}{=}\tilde{\bm{H}}_{k}^{\dag}},
\end{equation}
leading to the equivalence to the inequality in~(\ref{eq:psiJ}) as $ \tilde{\bm{\vartheta}}_R^{\dag} \tilde{\bm{H}}_{k}^{\dag} \bm{Z} \tilde{\bm{H}}_{k} \tilde{\bm{\vartheta}}_R \ge \psi_{Jk} $. Then, by introducing the trace operation and exploiting the properties of trace, we have $ \mathsf{Tr}\left( \tilde{\bm{H}}_{k}^{\dag} \bm{Z} \tilde{\bm{H}}_{k} \tilde{\bm{\vartheta}}_R \tilde{\bm{\vartheta}}_R^{\dag} \right) - \psi_{Jk} \ge 0 $. Further, by invoking the equality $ \mathsf{Tr}\left(\bm{A}^{\dag}\bm{B}\bm{C}\bm{D}\right) = \mathsf{vec}^{\dag}\left(\bm{A}\right)\left(\bm{D}^T\otimes \bm{B}\right) \mathsf{vec}\left(\bm{C}\right) $, we arrive at
\begin{equation} \label{eq:vecOrig}
	\mathsf{vec}^{\dag}\left( \tilde{\bm{H}}_{k} \right) \left( \left(\tilde{\bm{\vartheta}}_R \tilde{\bm{\vartheta}}_R^{\dag}\right)^T \otimes \bm{Z} \right) \mathsf{vec}\left( \tilde{\bm{H}}_{k} \right) - \psi_{Jk} \ge 0.
\end{equation}
Recall the channel uncertainties in~(\ref{eq:jkUnc}) and~(\ref{eq:jrkUnc}) with the definition in~(\ref{eq:psiJTilde}), we have that
\begin{equation}
	\tilde{\bm{H}}_{k} = \hat{\tilde{\bm{H}}}_{k} + \Delta\tilde{\bm{H}}_{k},
\end{equation}
where $ \hat{\tilde{\bm{H}}}_{k} = \left[ \hat{\bm{h}}_{Jk} \:\: \hat{\bm{H}}_{JRk}^{\dag} \right] $ and $ \Delta\tilde{\bm{H}}_{k} = \left[ \Delta\bm{h}_{Jk} \:\: \Delta\bm{H}_{JRk}^{\dag} \right] $. Then, the inequality in~(\ref{eq:vecOrig}) is extended as
\begin{equation} \label{eq:SPrim}
\begin{aligned}
	&\mathsf{vec}^{\dag}\left( \Delta\tilde{\bm{H}}_{k} \right) \bm{\Omega} \mathsf{vec}\left( \Delta\tilde{\bm{H}}_{k} \right) + \mathsf{vec}^{\dag}\left( \Delta\tilde{\bm{H}}_{k} \right) \bm{\Omega} \mathsf{vec}\left( \hat{\tilde{\bm{H}}}_{k} \right) \\
        & + \mathsf{vec}^{\dag}\left( \hat{\tilde{\bm{H}}}_{k} \right) \bm{\Omega} \mathsf{vec}\left( \Delta\tilde{\bm{H}}_{k} \right) + \mathsf{vec}^{\dag}\left( \hat{\tilde{\bm{H}}}_{k} \right) \bm{\Omega} \mathsf{vec}\left( \hat{\tilde{\bm{H}}}_{k} \right) \\
        & -\psi_{Jk} \ge 0,
\end{aligned}
\end{equation}
where $ \bm{\Omega} \overset{\Delta}{=} \left( \left(\tilde{\bm{\vartheta}}_R \tilde{\bm{\vartheta}}_R^{\dag}\right)^T \otimes \bm{Z} \right) $ is defined for notation simplicity. Meanwhile, for the uncertainty bounds in~(\ref{eq:jkUnc}), it can be rewritten as $ \Delta\bm{h}_{Jk}^{\dag}\Delta\bm{h}_{Jk} \le \epsilon_{Jk} $, in the equivalent form as
\begin{equation}
\mathsf{vec}^{\dag}\left(\Delta\bm{h}_{Jk}\right) \mathsf{vec}\left(\Delta\bm{h}_{Jk}\right) \le \epsilon_{Jk},	
\end{equation}
by applying the matrix equality $ \mathsf{Tr}\left( \bm{A}^{\dag}\bm{B} \right) = \mathsf{vec}^{\dag}\left(\bm{A}\right) \mathsf{vec}\left(\bm{B}\right) $. Comparing the structures of $ \Delta\bm{h}_{Jk} $ with $ \Delta\tilde{\bm{H}}_{k} $, we can rewrite the error bound for $ \Delta\bm{h}_{Jk} $ with respect to $ \Delta\tilde{\bm{H}}_{k} $ as
\begin{equation} \label{eq:SSec1}
\begin{aligned}
        &\mathsf{vec}^{\dag}\left( \Delta\tilde{\bm{H}}_{k} \right)
        \underbrace{
        \begin{bmatrix}
        \bm{1}_{M\times M} & \bm{0}_{M\times MN_R} \\
        \bm{0}_{MN_R \times M} & \bm{0}_{MN_R\times MN_R}
        \end{bmatrix}
        }_{\overset{\Delta}{=} \bm{\Upsilon}_{Jk} }
        \mathsf{vec}\left( \Delta\tilde{\bm{H}}_{k} \right) \\
        & - \epsilon_{Jk} \le 0,
\end{aligned}
\end{equation}
where $ \bm{\Upsilon}_{Jk} $ is introduced for notation simplicity. Similarly, for the error bounds in~(\ref{eq:jrkUnc}), we can derive the equivalence as
\begin{equation}
\mathsf{vec}^{\dag}\left(\Delta\bm{H}_{JRk}\right) \mathsf{vec}\left(\Delta\bm{H}_{JRk}\right) \le \epsilon_{JRk}.
\end{equation}
Then, the inequality above can be rewritten in terms of $ \Delta\tilde{\bm{H}}_{k} $ as
\begin{equation} \label{eq:SSec2}
\begin{aligned}
        &\mathsf{vec}^{\dag}\left( \Delta\tilde{\bm{H}}_{k} \right)
        \underbrace{
        \begin{bmatrix}
        \bm{0}_{M\times M} & \bm{0}_{M\times MN_R} \\
        \bm{0}_{MN_R \times M} & \bm{1}_{MN_R\times MN_R}
        \end{bmatrix}
        }_{\overset{\Delta}{=} \bm{\Upsilon}_{JRk} }
        \mathsf{vec}\left( \Delta\tilde{\bm{H}}_{k} \right) \\
        & - \epsilon_{JRk} \le 0,
\end{aligned}
\end{equation}
by comparing the elements in $ \Delta\bm{H}_{JRk} $ and $ \Delta\tilde{\bm{H}}_{k} $, with $ \bm{\Upsilon}_{JRk} $ similarly introduced. For the inequalities in~(\ref{eq:SPrim}),~(\ref{eq:SSec1}), and~(\ref{eq:SSec2}) with quadratic forms on the left-hand side, we can adopt general S-procedure~\cite{gens} to derive the following inequality in~(\ref{eq:SProc}),
\begin{figure*}
\begin{equation} \label{eq:SProc}
        \begin{bmatrix}
        \bm{\Omega} + \eta_{1,k}\bm{\Upsilon}_{Jk} + \eta_{2,k}\bm{\Upsilon}_{JRk} & \bm{\Omega} \mathsf{vec}\left( \hat{\tilde{\bm{H}}}_{k} \right) \\
        \mathsf{vec}^{\dag}\left( \hat{\tilde{\bm{H}}}_{k} \right)\bm{\Omega} & \mathsf{vec}^{\dag}\left( \hat{\tilde{\bm{H}}}_{k} \right) \bm{\Omega} \mathsf{vec}\left( \hat{\tilde{\bm{H}}}_{k} \right) - \psi_{Jk} - \eta_{1,k}\epsilon_{Jk} - \eta_{2,k}\epsilon_{JRk}
        \end{bmatrix}
        \succcurlyeq 0,
\end{equation}
\hrulefill
\end{figure*}
where $ \left\{\eta_{1,k}\right\}_{k\in\mathcal{K}} $ and $ \left\{\eta_{2,k}\right\}_{k\in\mathcal{K}} $ are the introduced non-negative variables associated with the condition in~(\ref{eq:SSec1}) and~(\ref{eq:SSec2}) in the general S-procedure, respectively.

With previous operations of introducing the auxiliary variables in~(\ref{eq:psiS}) and~(\ref{eq:psiJ}), along with the reformulation against the uncertainties resulting in~(\ref{eq:signDef}) and~(\ref{eq:SProc}), we reach a deterministic problem eliminating the uncertainties as a lower bound for the original inner optimization in~(\ref{eq:innerProb}), specified as
\begin{IEEEeqnarray}{cl} \hspace{-10pt}
	\IEEEyesnumber\label{eq:innerProbIter} \IEEEyessubnumber*
        \max_{\substack{\bm{\vartheta}_A,\bm{\vartheta}_R, \bm{Z}, \\ \left\{\psi_{Sk},\psi_{Jk}\right\}_{k\in\mathcal{K}} \\ \left\{\rho_{1,k},\rho_{2,k}\right\}_{k\in\mathcal{K}}, \\ \left\{\eta_{1,k},\eta_{2,k}\right\}_{k\in\mathcal{K}}}} \:\: & \begin{aligned} &\log\left(1 + \gamma_D\right) \\ & - \max_{k\in\mathcal{K}} \log\left(1 + \frac{P_S \psi_{Sk}}{\psi_{Jk} + \sigma_0^2}\right) \end{aligned}
        \\
	 {\:\:\:\rm{s.t.}} \quad & (\ref{eq:inAConstr}), (\ref{eq:inRConstr}), (\ref{eq:inzConstr}), \\
	 & (\ref{eq:signDef}), (\ref{eq:SProc}), \quad\forall k\in\mathcal{K}, \label{eq:uncConstr} \\
	 & \rho_{1,k}, \rho_{2,k}\ge 0, \eta_{1,k}, \eta_{2,k}\ge 0, \forall k\in\mathcal{K}. \label{eq:auxConstr}
\end{IEEEeqnarray}
For the problems in~(\ref{eq:innerProb}) and~(\ref{eq:innerProbIter}), the new constraints in~(\ref{eq:uncConstr}) and~(\ref{eq:auxConstr}) combat the uncertainties with the corresponding error bounds, and thus the solution to~(\ref{eq:innerProbIter}) achieves robustness as compared with the original counterpart in~(\ref{eq:innerProb}). Furthermore, we introduce a new variable $ \varphi $, to tackle the non-continuous operation in the objective function, and induce the optimization problem as
\begin{IEEEeqnarray}{cl} \hspace{-15pt}
	\IEEEyesnumber\label{eq:innerProbFin} \IEEEyessubnumber*
	\max_{\substack{\bm{\vartheta}_A,\bm{\vartheta}_R, \bm{Z},\varphi, \\ \left\{\psi_{Sk},\psi_{Jk}\right\}_{k\in\mathcal{K}},  \\ \left\{\rho_{1,k},\rho_{2,k}\right\}_{k\in\mathcal{K}}, \\ \left\{\eta_{1,k},\eta_{2,k}\right\}_{k\in\mathcal{K}}}} \:\: & \log\left(1 + \gamma_D\right) - \varphi \\
	 {\:\:\:\rm{s.t.}} \quad & (\ref{eq:inAConstr}), (\ref{eq:inRConstr}), (\ref{eq:inzConstr}), \\
	 & (\ref{eq:signDef}), (\ref{eq:SProc}), \quad\forall k\in\mathcal{K}, \\
	 & \rho_{1,k}, \rho_{2,k}\ge 0, \eta_{1,k}, \eta_{2,k}\ge 0, \forall k\in\mathcal{K}, \\
	 & \log\left(1 + \frac{P_S \psi_{Sk}}{\psi_{Jk} + \sigma_0^2}\right) \le \varphi, \forall k\in\mathcal{K}, \\
	 & \varphi \ge 0,
\end{IEEEeqnarray}
which facilitates further discussions to solve for jamming and reflection strategies as detailed below.

\subsection{Jamming Optimization}

We first address the jamming subproblem while considering fixed reflection coefficients at the RISs, i.e., to solve for $ \bm{Z} $ with fixed $ \bm{\vartheta}_A $ and $ \bm{\vartheta}_R $ in~(\ref{eq:innerProbFin}). Correspondingly, we ignore the constraints related to the reflection coefficients and simplify the problem as
\begin{IEEEeqnarray}{cl} \hspace{-15pt}
	\IEEEyesnumber\label{eq:innerProbZ} \IEEEyessubnumber*
	\max_{\substack{ \bm{Z}, \left\{\psi_{Sk},\psi_{Jk}\right\}_{k\in\mathcal{K}}, \\ \varphi, \left\{\rho_{1,k},\rho_{2,k}\right\}_{k\in\mathcal{K}}, \\ \left\{\eta_{1,k},\eta_{2,k}\right\}_{k\in\mathcal{K}}}} \:\: & \log\left(1 + \gamma_D \right) - \varphi \\
	 {\:\:\:\rm{s.t.}} \quad &  (\ref{eq:inzConstr}), \label{eq:pwrZZ}\\
	 & (\ref{eq:signDef}), (\ref{eq:SProc}), \quad\forall k\in\mathcal{K}, \label{eq:robustZ} \\
	 & \rho_{1,k}, \rho_{2,k}\ge 0, \eta_{1,k}, \eta_{2,k}\ge 0, \forall k\in\mathcal{K}, \label{eq:auxZ}\\
	 & \log\left(1 + \frac{P_S \psi_{Sk}}{\psi_{Jk} + \sigma_0^2}\right) \le \varphi, \forall k\in\mathcal{K}, \label{eq:varphiConZ}\\
	 & \varphi \ge 0. \label{eq:varphiZ}
\end{IEEEeqnarray}
As the jamming optimization variable is incorporated in $ \gamma_D $, we can rewrite $ \gamma_D $ as
\begin{equation}
	\gamma_D = \frac{P_{SD}}{ \tilde{\bm{h}}_{JD}^{\dag}\bm{Z}\tilde{\bm{h}}_{JD} + \sigma_0^2 },
\end{equation}
where
\begin{equation}
\begin{aligned}
P_{SD} \overset{\Delta}{=}& P_S \left| \bm{\vartheta}_A^{\dag} \bm{h}_{SAD} + \bm{\vartheta}_R^{\dag} \bm{h}_{SRD} \right|^2, \\
\tilde{\bm{h}}_{JD}^{\dag} \overset{\Delta}{=}& \left( \bm{h}_{Jk}^{\dag} + \bm{\vartheta}_R^{\dag} \bm{H}_{JRk} \right),
\end{aligned}
\end{equation}
are defined for notation simplicity. By substituting the equality $ \tilde{\bm{h}}_{JD}^{\dag}\bm{Z}\tilde{\bm{h}}_{JD} = \mathsf{Tr}\left(\bm{Z}\tilde{\bm{H}}_{JD}\right) $ with $ \tilde{\bm{H}}_{JD} = \tilde{\bm{h}}_{JD}^{\dag}\tilde{\bm{h}}_{JD} $ into the objective function, we can see that the problem in~(\ref{eq:innerProbZ}) is a semidefinite programming (SDP) problem with respect to jamming optimization. Also, the non-convexity in~(\ref{eq:innerProbZ}) lies in the objective function and the constraint in~(\ref{eq:varphiConZ}). In this regard, by applying~\cite[Lemma~1]{lem}, we introduce an auxiliary variable $ t_{JD} $, to linearize the non-concave term to approximate the objective function as
\begin{equation} \label{eq:RSJ}
\begin{aligned}
        \bar{R}_S = &\log\left( P_{SD} + \mathsf{Tr}\left(\bm{Z}\tilde{\bm{H}}_{JD}\right) + \sigma_0^2 \right) \\
        &- t_{JD} \left( \mathsf{Tr}\left(\bm{Z}\tilde{\bm{H}}_{JD}\right) + \sigma_0^2 \right) + \log t_{JD} + 1 - \varphi,  
\end{aligned}
\end{equation}
which amounts to the original objective function on condition that
\begin{equation} \label{eq:tJD}
	t_{JD} = \left( \mathsf{Tr}\left(\bm{Z}\tilde{\bm{H}}_{JD}\right) + \sigma_0^2 \right)^{-1}.
\end{equation}
For the non-convex constraint in~(\ref{eq:varphiConZ}), we can employ the same procedure above to approximate it as
\begin{equation} \label{eq:tkLem}
	-t_{k}\left( P_S\psi_{Sk} + \psi_{Jk} + \sigma_0^2 \right) + \log t_{k} + 1 + \log\left(\psi_{Jk} + \sigma_0^2\right) + \varphi \ge 0,
\end{equation}
which equals the original when
\begin{equation} \label{eq:tk}
	t_{k} = \left( P_S\psi_{Sk} + \psi_{Jk} + \sigma_0^2 \right)^{-1}.
\end{equation}

Through the operations above, we reformulate the problem in~(\ref{eq:innerProbZ}) as
\begin{IEEEeqnarray}{cl}
	\IEEEyesnumber\label{eq:innerProbZRefm} \IEEEyessubnumber*
	\max_{\substack{ \bm{Z}, \left\{\psi_{Sk},\psi_{Jk}\right\}_{k\in\mathcal{K}}, \varphi, \\ \left\{\rho_{1,k},\rho_{2,k}\right\}_{k\in\mathcal{K}}, \\ \left\{\eta_{1,k},\eta_{2,k}\right\}_{k\in\mathcal{K}}}} \:\: & \bar{R}_S \\
	 {\:\:\:\rm{s.t.}} \quad &  (\ref{eq:pwrZZ}),(\ref{eq:robustZ}),(\ref{eq:auxZ}),(\ref{eq:varphiZ}), \\
	 & (\ref{eq:tkLem}), \quad\forall k\in\mathcal{K}.
\end{IEEEeqnarray}
The problem in~(\ref{eq:innerProbZRefm}) can be easily verified as a convex optimization with respect to the optimization variables, and thus can be conveniently solved with off-the-shelf solvers. Then, the optimum through~(\ref{eq:innerProbZRefm}) needs to be substituted into~(\ref{eq:tJD}) and~(\ref{eq:tk}) to update the auxiliary variables. Finally, the problem solving in~(\ref{eq:innerProbZRefm}) and updates in~(\ref{eq:tJD}) and~(\ref{eq:tk}) are conducted in an iterative manner, where the convergence brings the optimal jamming beamforming for secure transmissions.

\subsection{Reflection Optimization at the ARIS}

Then, we consider the reflection optimization at the ARIS, while treating jamming and reflection at the fixed RIS as constants, i.e., to solve for $ \bm{\vartheta}_A $ with fixed $ \bm{Z} $ and $ \bm{\vartheta}_R $ in~(\ref{eq:innerProbFin}). In this regard, the problem is simplified as
\begin{IEEEeqnarray}{cl} \hspace{-18pt}
	\IEEEyesnumber\label{eq:innerProbFinA} \IEEEyessubnumber*
	\max_{\substack{\bm{\vartheta}_A, \left\{\psi_{Sk},\psi_{Jk}\right\}_{k\in\mathcal{K}}, \\ \varphi,  \left\{\rho_{1,k},\rho_{2,k}\right\}_{k\in\mathcal{K}},\\ \left\{\eta_{1,k},\eta_{2,k}\right\}_{k\in\mathcal{K}}}} \:\: & \log\left(1 + \gamma_D\right) - \varphi \\
	 {\:\:\:\rm{s.t.}} \quad & (\ref{eq:inAConstr}),  \\
	 & (\ref{eq:signDef}), (\ref{eq:SProc}), \quad\forall k\in\mathcal{K}, \label{eq:robustA} \\
	 & \rho_{1,k}, \rho_{2,k}\ge 0, \eta_{1,k}, \eta_{2,k}\ge 0, \forall k\in\mathcal{K}, \label{eq:auxA}\\
	 & \log\left(1 + \frac{P_S \psi_{Sk}}{\psi_{Jk} + \sigma_0^2}\right) \le \varphi, \forall k\in\mathcal{K}, \label{eq:varphiConA} \\
	 & \varphi \ge 0. \label{eq:varphiA}
\end{IEEEeqnarray}
To facilitate the problem solving, we rewrite $ \gamma_D $ as
\begin{equation} \label{eq:gammaDA}
	\gamma_D = \frac{P_S}{P_{JD} + \sigma_0^2} \left| \bm{\vartheta}_A^{\dag} \bm{h}_{SAD} + {g}_{SRD} \right|^2,
\end{equation}
where
\begin{equation}
\begin{aligned}
        P_{JD} \overset{\Delta}{=}& \left( \bm{h}_{JD}^{\dag} + \bm{\vartheta}_R^{\dag} \bm{h}_{JRD} \right) \bm{Z} \left( \bm{h}_{JD}^{\dag} + \bm{\vartheta}_R^{\dag} \bm{h}_{JRD} \right)^{\dag}, \\
        {g}_{SRD} \overset{\Delta}{=}& \bm{\vartheta}_R^{\dag} \bm{h}_{SRD},        
\end{aligned}
\end{equation}
are defined for notation simplicity. As we can see in~(\ref{eq:gammaDA}) that $ \gamma_D $ is a quadratic and thus convex function with respect to $ \bm{\vartheta}_A $, we can then exploit the first-order approximation as a lower-bound at $ \bm{\vartheta}_A^{\circ} $ to linearize it as
\begin{equation} \label{eq:gammaDApp}
	\gamma_D \ge \frac{P_S}{P_{JD} + \sigma_0^2} \Phi_A\left(\bm{\vartheta}_A; \bm{\vartheta}_A^{\circ}\right),
\end{equation}
where
\begin{equation}
\begin{aligned}
	\Phi_A\left(\bm{\vartheta}_A; \bm{\vartheta}_A^{\circ}\right) = &{\:} \bm{h}_{SAD}^{\dag} \bm{\vartheta}_A^{\circ} \bm{\vartheta}_A^{\dag} \bm{h}_{SAD} + \bm{\vartheta}_A^{\dag} \bm{h}_{SAD} g_{SRD}^{\dag} \\
	&{} + \bm{h}_{SAD}^{\dag} \bm{\vartheta}_A \left(\bm{\vartheta}_A^{\circ}\right)^{\dag} \bm{h}_{SAD} + g_{SRD} \bm{h}_{SAD}^{\dag} \bm{\vartheta}_A \\
	&{} - \bm{h}_{SAD}^{\dag} \bm{\vartheta}_A^{\circ} \left(\bm{\vartheta}_A^{\circ}\right)^{\dag} \bm{h}_{SAD} + g_{SRD}^{\dag}g_{SRD}.
\end{aligned}
\end{equation}
Then, for the unit-modulus constraint regarding the elements of $ \bm{\vartheta}_A $, we can convert it into the joint constraints as
\begin{equation} \label{eq:ccpA}
	\left| \vartheta_{A,n} \right|^2 \ge 1, \quad \left| \vartheta_{A,n} \right|^2 \le 1, \quad\forall n\in\mathcal{N}_A,
\end{equation}
where the first inequality with non-convexity can be linearly approximated at $ \bm{\vartheta}_A^{\circ} $ as
\begin{equation} \label{eq:ccpAApp}
	2\mathsf{Re}\left\{ \left(\vartheta_{A,n}^{\circ}\right)^{\dag}\vartheta_{A,n} \right\} - \left| \vartheta_{A,n}^{\circ} \right|^2 + 1 \le 0, \quad\forall n\in\mathcal{N}_A.
\end{equation}
Also, the non-convex constraint in~(\ref{eq:varphiConA}) can be similarly treated as that in~(\ref{eq:tkLem}), with the auxiliary variable $ t_k $ that equalizes the original constraint in~(\ref{eq:varphiConA}) when the condition in~(\ref{eq:tk}) is satisfied.

Based on the discussions above, we arrive at a convex counterpart of the ARIS reflection problem in~(\ref{eq:innerProbFinA}), given as
\begin{IEEEeqnarray}{cl} \hspace{-10pt}
	\IEEEyesnumber\label{eq:innerProbFinARefm} \IEEEyessubnumber*
	\max_{\substack{\bm{\vartheta}_A, \left\{\psi_{Sk},\psi_{Jk}\right\}_{k\in\mathcal{K}},\\ \varphi, \left\{\rho_{1,k},\rho_{2,k}\right\}_{k\in\mathcal{K}}, \\ \left\{\eta_{1,k},\eta_{2,k}\right\}_{k\in\mathcal{K}}, \\ \left\{ \iota_{A,n} \right\}_{n=1,2,\cdots,2N_A} }} \:\: & \begin{aligned} &\log\left(1 + \frac{P_S}{P_{JD} + \sigma_0^2} \Phi_A\left(\bm{\vartheta}_A; \bm{\vartheta}_A^{\circ}\right) \right) \\ & - \varphi - \lambda_A\sum\limits_{n=1}^{2N_A}\iota_{A,n} \end{aligned} \\
	 {\:\:\:\rm{s.t.}} \quad & (\ref{eq:robustA}), (\ref{eq:auxA}),(\ref{eq:varphiA}), \\
	 & (\ref{eq:tkLem}), \quad \forall k\in\mathcal{K}, \\
	 & \left| \vartheta_{A,n} \right|^2 \le 1 + \iota_{A,n}, \quad\forall n\in\mathcal{N}_A \\
	 & \begin{aligned} & 2\mathsf{Re}\left\{ \left(\vartheta_{A,n}^{\circ}\right)^{\dag}\vartheta_{A,n} \right\} - \left| \vartheta_{A,n}^{\circ} \right|^2 \\ &\le -1 + \iota_{A,N_A+n} , \:\:\forall n\in\mathcal{N}_A, \end{aligned}
\end{IEEEeqnarray}
which is an approximation at $ \bm{\vartheta}_A^{\circ} $, and $ \left\{\iota_{A,n}\right\}_{n=1,2\cdots,2N_A} $ are additionally introduced to improve the convergence with $ \lambda_A $ as the coefficient for penalty. Then, the problem solving of~(\ref{eq:innerProbFinARefm}) for ARIS reflection and the auxiliary variable update in~(\ref{eq:tk}) are conducted in an iterative manner to obtain the current optimum, denoted by $ \bm{\vartheta}_A^{\star} $, on condition of the approximation at $ \bm{\vartheta}_A^{\circ} $. Finally, a successive convex approximation (SCA) procedure is conducted that the approximation point is updated with the current optimum to reach the next-round optimal reflection, and the convergence of the SCA procedure provides the optimal ARIS reflection coefficients.

\subsection{Reflection Optimization at the Fixed RIS}

Considering constant jamming and reflection at the ARIS, the reflection optimization at the fixed RIS is given as
\begin{IEEEeqnarray}{cl} \hspace{-20pt}
	\IEEEyesnumber\label{eq:innerProbFinR} \IEEEyessubnumber*
	\max_{\substack{\bm{\vartheta}_R, \left\{\psi_{Sk},\psi_{Jk}\right\}_{k\in\mathcal{K}}, \\ \varphi, \left\{\rho_{1,k},\rho_{2,k}\right\}_{k\in\mathcal{K}}, \\ \left\{\eta_{1,k},\eta_{2,k}\right\}_{k\in\mathcal{K}} }} \:\: & \log\left(1 + \gamma_D\right) - \varphi \\
	 {\:\:\:\rm{s.t.}} \quad & (\ref{eq:inRConstr}),  \\
	 & (\ref{eq:signDef}), (\ref{eq:SProc}), \quad\forall k\in\mathcal{K}, \label{eq:robustR} \\
	 & \rho_{1,k}, \rho_{2,k}\ge 0, \eta_{1,k}, \eta_{2,k}\ge 0, \forall k\in\mathcal{K}, \label{eq:auxR}\\
	 & \log\left(1 + \frac{P_S \psi_{Sk}}{\psi_{Jk} + \sigma_0^2}\right) \le \varphi, \forall k\in\mathcal{K}, \label{eq:varphiConR} \\
	 & \varphi \ge 0. \label{eq:varphiR}
\end{IEEEeqnarray}
Given the complicated relationship between the SINR at the legitimate receiver and considered reflection coefficient, we introduce a new variable $ \psi_{JD} $ and reformulate $ \gamma_D $ as
\begin{equation} \label{eq:gammaDAppLB}
	\gamma_D \ge \frac{P_S}{\psi_{JD} + \sigma_0^2} \left| g_{SAD} + \bm{\vartheta}_R^{\dag} \bm{h}_{SRD} \right|^2,
\end{equation}
where
\begin{equation} \label{eq:psiJD}
\begin{aligned}
       \psi_{JD} \ge& \left( \bm{h}_{JD}^{\dag} + \bm{\vartheta}_R^{\dag} \bm{h}_{JRD} \right) \bm{Z} \left( \bm{h}_{JD}^{\dag} + \bm{\vartheta}_R^{\dag} \bm{h}_{JRD} \right)^{\dag}, \\
       g_{SAD} \overset{\Delta}{=}& \bm{\vartheta}_A^{\dag} \bm{h}_{SAD}.       
\end{aligned}
\end{equation}
For the inequality regarding $ \psi_{JD} $ in~(\ref{eq:psiJD}), we can adopt the Schur complement to recast it in the form of linear matrix inequality as
\begin{equation} \label{eq:schur}
	\begin{bmatrix}
		\psi_{JD} & \bm{h}_{JD}^{\dag} + \bm{\vartheta}_R^{\dag} \bm{h}_{JRD} \\
		\left( \bm{h}_{JD}^{\dag} + \bm{\vartheta}_R^{\dag} \bm{h}_{JRD} \right)^{\dag} & \bm{Z}
	\end{bmatrix}
	\succcurlyeq 0.
\end{equation}
Further, for the quadratic term with respect to reflection coefficient in~(\ref{eq:gammaDAppLB}), we can employ the first-order Taylor expansion similarly as~(\ref{eq:gammaDApp}) to reach that
\begin{equation} \label{eq:gammaDLB}
	\gamma_D \ge \frac{P_S}{\psi_{JD} + \sigma_0^2} \Phi_R\left(\bm{\vartheta}_R; \bm{\vartheta}_R^{\circ}\right),
\end{equation}
as an approximation at $ \bm{\vartheta}_R^{\circ} $ with
\begin{equation}
\begin{aligned}
	\Phi_R\left(\bm{\vartheta}_R; \bm{\vartheta}_R^{\circ}\right) = &{\:} \bm{h}_{SRD}^{\dag} \bm{\vartheta}_R^{\circ} \bm{\vartheta}_R^{\dag} \bm{h}_{SRD} + \bm{\vartheta}_R^{\dag} \bm{h}_{SRD} g_{SAD}^{\dag} \\
	&{} + \bm{h}_{SRD}^{\dag} \bm{\vartheta}_R \left(\bm{\vartheta}_R^{\circ}\right)^{\dag} \bm{h}_{SRD} + g_{SAD} \bm{h}_{SRD}^{\dag} \bm{\vartheta}_R \\
	&{} - \bm{h}_{SRD}^{\dag} \bm{\vartheta}_R^{\circ} \left(\bm{\vartheta}_R^{\circ}\right)^{\dag} \bm{h}_{SRD} + g_{SAD}^{\dag}g_{SAD}.
\end{aligned}
\end{equation}
By replacing $ \gamma_D $ with the lower bound given in~(\ref{eq:gammaDLB}), the objective function in~(\ref{eq:innerProbFinR}) is now concave with respect to the reflection coefficients. When jointly considering the newly introduced variable $ \psi_{JD} $, we can adopt the same technique as~(\ref{eq:RSJ}) to reformulate the objective function as
\begin{equation}
\begin{aligned}
        \bar{\bar{R}}_S =& \log\left( P_S \Phi_R + \psi_{JD} + \sigma_0^2 \right) - t_{RD} \left( \psi_{JD} + \sigma_0^2 \right) \\
        &+ \log t_{JD} + 1 - \varphi,        
\end{aligned}
\end{equation}
with an introduced variable $ t_{RD} $. Also similar as before, the reformulation is equivalent on condition that
\begin{equation} \label{eq:tRD}
	t_{RD} = \left( \psi_{JD} + \sigma_0^2 \right)^{-1}.
\end{equation}

Then, the unit-modulus constraint regarding the elements of $ \bm{\vartheta}_R $ can be similarly treated as~(\ref{eq:ccpA}) and~(\ref{eq:ccpAApp}), leading to
\begin{equation}
	\left| \vartheta_{R,n} \right|^2 \le 1,  2\mathsf{Re}\left\{ \left(\vartheta_{R,n}^{\circ}\right)^{\dag}\vartheta_{R,n} \right\} - \left| \vartheta_{R,n}^{\circ} \right|^2 + 1 \le 0, \forall n\in\mathcal{N}_R,
\end{equation}
approximated at $ \bm{\vartheta}_R^{\circ} $. Meanwhile, the constraint in~(\ref{eq:SProc}) is no longer a linear matrix inequality with respect to $ \bm{\vartheta}_R $. For this issue, the non-linear part traces back to the inequality in~(\ref{eq:vecOrig}) incorporating the quadratic term against the reflection coefficients. Recalling that $ \tilde{\bm{\vartheta}}^{\dag} = \left[ 1 \:\: \bm{\vartheta}_R^{\dag} \right] $, we can use the first-order approximation at $ \bm{\vartheta}_R^{\circ} $ given as
\begin{equation}
\begin{aligned}
        \tilde{\bm{\vartheta}}_R \tilde{\bm{\vartheta}}_R^{\dag} \ge &
        \begin{bmatrix}
                1 & \bm{\vartheta}_R^{\dag} \\
                \bm{\vartheta}_R^{\circ} & \bm{\vartheta}_R^{\circ}\bm{\vartheta}_R^{\dag}
        \end{bmatrix}
        +
        \begin{bmatrix}
                1 & \left(\bm{\vartheta}_R^{\circ}\right)^{\dag} \\
                \bm{\vartheta}_R & \bm{\vartheta}_R\left(\bm{\vartheta}_R^{\circ}\right)^{\dag}
        \end{bmatrix} \\
        &-
        \begin{bmatrix}
                1 & \left(\bm{\vartheta}_R^{\circ}\right)^{\dag} \\
                \bm{\vartheta}_R^{\circ} & \bm{\vartheta}_R^{\circ}\left(\bm{\vartheta}_R^{\circ}\right)^{\dag}
        \end{bmatrix}
        \overset{\Delta}{=} {\bm{\Psi}}_R\left(\bm{\vartheta}_R; \bm{\vartheta}_R^{\circ}\right).        
\end{aligned}
\end{equation}
Correspondingly, by defining $ \bm{\Xi} \overset{\Delta}{=} {\bm{\Psi}}^T_R\left(\bm{\vartheta}_R; \bm{\vartheta}_R^{\circ}\right) \otimes \bm{Z} $, we have $ \bm{\Omega} \ge \bm{\Xi} $, and the inequality in~(\ref{eq:vecOrig}) is approximated as
\begin{equation}
	\mathsf{vec}^{\dag}\left( \tilde{\bm{H}}_{k} \right) \bm{\Xi} \mathsf{vec}\left( \tilde{\bm{H}}_{k} \right) - \psi_{Jk} \ge 0.
\end{equation}
Then, with the general S-Procedure conducted similarly as that in Sec.~\ref{sec:IV-A}, we reach an approximated version of the inequality in~(\ref{eq:SProc}) at $ \bm{\vartheta}_R^{\circ} $, given as~(\ref{eq:SProcApp}),
\begin{figure*}[t]
\begin{equation} \label{eq:SProcApp}
	\begin{bmatrix}
	\bm{\Xi} + \eta_{1,k}\bm{\Upsilon}_{Jk} + \eta_{2,k}\bm{\Upsilon}_{JRk} & \bm{\Xi} \mathsf{vec}\left( \hat{\tilde{\bm{H}}}_{k} \right) \\
	\mathsf{vec}^{\dag}\left( \hat{\tilde{\bm{H}}}_{k} \right)\bm{\Xi} & \mathsf{vec}^{\dag}\left( \hat{\tilde{\bm{H}}}_{k} \right) \bm{\Xi} \mathsf{vec}\left( \hat{\tilde{\bm{H}}}_{k} \right) - \psi_{Jk} - \eta_{1,k}\epsilon_{Jk} - \eta_{2,k}\epsilon_{JRk}
	\end{bmatrix}
	\succcurlyeq 0,
\end{equation}
\hrulefill
\end{figure*}
which is a linear matrix inequality with respect to $ \bm{\vartheta}_R $. Finally, the non-convex constraint in~(\ref{eq:varphiConR}) can be similarly tackled as~(\ref{eq:tkLem}) and~(\ref{eq:tk}).

With the reformulations above, we arrive at an convex problem given as
\begin{IEEEeqnarray}{cl} \hspace{-10pt}
	\IEEEyesnumber\label{eq:innerProbFinRRefm} \IEEEyessubnumber*
	\max_{\substack{\bm{\vartheta}_R, \left\{\psi_{Sk},\psi_{Jk}\right\}_{k\in\mathcal{K}}, \\ \varphi, \left\{\rho_{1,k},\rho_{2,k}\right\}_{k\in\mathcal{K}}, \\ \left\{\eta_{1,k},\eta_{2,k}\right\}_{k\in\mathcal{K}}, \\ \psi_{JD}, \left\{ \iota_{R,n} \right\}_{n=1,2,\cdots,2N_R} }} \:\: & \bar{\bar{R}}_S - \lambda_R\sum\limits_{n=1}^{2N_R}\iota_{R,n} \\
	 {\:\:\:\rm{s.t.}} \quad & (\ref{eq:signDef}), (\ref{eq:tkLem}), (\ref{eq:SProcApp}), \quad \forall k\in\mathcal{K}, \\
	 & (\ref{eq:varphiR}), (\ref{eq:schur}),  \\
	 & \left| \vartheta_{R,n} \right|^2 \le 1 + \iota_{R,n}, \forall n\in\mathcal{N}_R, \\
	 & \begin{aligned} & 2\mathsf{Re}\left\{ \left(\vartheta_{R,n}^{\circ}\right)^{\dag}\vartheta_{R,n} \right\} - \left| \vartheta_{R,n}^{\circ} \right|^2 \\ & \le -1 + \iota_{R,N_R+n} , \:\:\forall n\in\mathcal{N}_R, \end{aligned}
\end{IEEEeqnarray}
which is approximated at $ \bm{\vartheta}^{\circ} $, and similarly as in~(\ref{eq:innerProbFinARefm}), the variables $ \left\{\iota_{R,n}\right\}_{n=1,2\cdots,2N_R} $ are introduced to improve the convergence. As we solve the problem in~(\ref{eq:innerProbFinRRefm}) to obtain the reflection coefficients, the auxiliary parameters are updated according to~(\ref{eq:tk}) and~(\ref{eq:tRD}), and this process is continued until the convergence brings the current optimum, denoted by $ \bm{\vartheta}_R^{\star} $. Then, we employ the SCA technique to use the current optimum as the next-round approximation point, i.e., $ \bm{\vartheta}_R^{\circ} \gets \bm{\vartheta}_R^{\star} $, to further update the reflection coefficients. The convergence of the SCA procedure brings the optimal reflection at the fixed RIS.

\begin{algorithm}[t] \label{alg:1} \scriptsize
  \caption{BCD framework for secure transmission}
  Initialization: $ \tau \gets 0 $;  randomly select jamming and reflection strategies satisfying the constraints in~(\ref{eq:innerProb}), denoted as $ \bm{Z}^{(\tau)} $, $ \bm{\vartheta}_A^{(\tau)} $, $ \bm{\vartheta}_R^{(\tau)} $\;
  Reformulate the problem as~(\ref{eq:innerProbFin}) eliminating the uncertainties\;
  \Repeat(\tcp*[f]{BCD procedure to update jamming and reflection}){\scriptsize $ \left| \left[ \mathsf{vec}\left(\bm{Z}^{(\tau)}\right); \bm{\vartheta}_A^{(\tau)}; \bm{\vartheta}_R^{(\tau)} \right] - \left[ \mathsf{vec}\left(\bm{Z}^{(\tau-1)}\right); \bm{\vartheta}_A^{(\tau-1)}; \bm{\vartheta}_R^{(\tau-1)} \right] \right| < \varepsilon $ } 
  {
  	$ \tau \gets \tau + 1 $; $ \bm{\vartheta}_A^{\star} \gets \bm{\vartheta}_A^{(\tau-1)} $; $ \bm{\vartheta}_R^{\star} \gets \bm{\vartheta}_R^{(\tau-1)} $\;
  	Construct the problem in~(\ref{eq:innerProbZRefm}) with $ \bm{\vartheta}_A^{(\tau-1)} $, $ \bm{\vartheta}_R^{(\tau-1)} $ \tcp*{SDP for jamming optimization}
  	Solve the problem in~(\ref{eq:innerProbZRefm}) and update $ \left\{t_k\right\}_{k\in\mathcal{K}} $ according to~(\ref{eq:tk}) iteratively, and obtain $ \bm{Z}^{(\tau)} $ at the convergence\;
    \Repeat(\tcp*[f]{SCA to update reflection at the ARIS}){$ \left| \bm{\vartheta}_A^{\circ} - \bm{\vartheta}_A^{\star} \right| \le \varepsilon_A $}
    {
    	$ \bm{\vartheta}_A^{\circ} \gets \bm{\vartheta}_A^{\star} $\;
    	Construct the problem in~(\ref{eq:innerProbFinARefm}) with $ \bm{Z}^{(\tau)} $, $ \bm{\vartheta}_R^{(\tau-1)} $ with approximation point $ \bm{\vartheta}_A^{\circ} $ \;
    	Iteratively solve the problem in~(\ref{eq:innerProbFinARefm}) and update $ \left\{t_k\right\}_{k\in\mathcal{K}} $ according to~(\ref{eq:tk}), and obtain $ \bm{\vartheta}_A^{\star} $ at the convergence\;
    }
    $ \bm{\vartheta}_A^{(\tau)} \gets \bm{\vartheta}_A^{\star} $\;
    \Repeat(\tcp*[f]{SCA to update reflection at the fixed RIS}){$ \left| \bm{\vartheta}_R^{\circ} - \bm{\vartheta}_R^{\star} \right| \le \varepsilon_R $}
    {
    	$ \bm{\vartheta}_R^{\circ} \gets \bm{\vartheta}_R^{\star} $\;
    	Construct the problem in~(\ref{eq:innerProbFinRRefm}) with $ \bm{Z}^{(\tau)} $, $ \bm{\vartheta}_A^{(\tau)} $ with approximation point $ \bm{\vartheta}_R^{\circ} $ \;
    	Iteratively solve the problem in~(\ref{eq:innerProbFinRRefm}) and update $ \left\{t_k\right\}_{k\in\mathcal{K}} $ and $ t_{RD} $ according to~(\ref{eq:tk}) and~(\ref{eq:tRD}), respectively, and obtain $ \bm{\vartheta}_R^{\star} $ at the convergence\;
    }
    $ \bm{\vartheta}_R^{(\tau)} \gets \bm{\vartheta}_R^{\star} $\;
  }
\end{algorithm}

\subsection{Algorithm Design}

In the preceding discussions, we have tackled the uncertainties to formulate the robust secrecy optimization problem, where the jamming and reflection optimizations are analyzed separately. Then, we can employ the BCD framework to update the jamming beamforming, ARIS reflection, and the reflection at the fixed RIS in an iterative manner, and the convergence achieves a suboptimum towards the robust secrecy optimization. The algorithm is summarized in Alg.~\ref{alg:1}, where the outer loop is for BCD framework, where $ \tau = 0,1,2,\cdots $ indicates the iterations and the constant $ \varepsilon $ claims the convergence. Meanwhile, obtaining the reflection at the RISs requires inner loops in the form of SCA procedures, where the constants $ \varepsilon_A $ and $ \varepsilon_R $ indicate the convergence.

\section{Learning for Deployment} \label{sec5}

In this section, we consider the deployment issue of the aerial platform as the outer subproblem of the original optimization in~(\ref{eq:problem}). The deployment affects the wireless channels related to the aerial platform, and further impacts the aerial reflection and cooperative jamming. Given the double-layer structure to solve the problem, the deployment in the outer layer is evaluated with inner problem providing the intermediate results. In this regard, we adopt the deep reinforcement learning technique to determine the deployment as the outer problem. The reinforcement learning technique enables effective decision-making due to its goal-driven nature while adapting to the environment, which has been widely used in existing researches~\cite{T}. For our considered problem, as the inner problem can be efficiently solved through optimization as detailed before, the resultant secrecy performance then significantly facilitates the learning process towards the optimal deployment.

\subsection{MDP Formulation}

The learning-based solution to determine the deployment can be formulated as a Markov decision process (MDP). A MDP is considered over a time series denoted by $ \mathcal{T} = \left\{0,1,\cdots, t, \cdots, T \right\} $, along with the state space, action space, and reward. For our considered problem of robust secrecy optimization, these components are elaborated as follows.
\begin{itemize}
	\item State space: For the considered problem, the robust secrecy rate inherently depends on the network topology and the bound of channel imperfection.
	As such, given fixed locations of source, destination, and fixed RIS, the state is defined as the set consisting of the location of the aerial platform, channel condition in the network, and the associated uncertainties, given as
	\begin{equation}
	\begin{aligned}
		\bm{s}\left(t\right) = \left\{ \bm{h}_{SAD}, \bm{h}_{SRD}, \bm{h}_{JD}, \bm{h}_{JRD}, \right. \\
                \left\{\hat{\bm{h}}_{SAk}, \hat{\bm{h}}_{SRk}, \hat{\bm{h}}_{Jk}, \hat{\bm{H}}_{JRk}\right\}_{k\in\mathcal{K}}, \\
		\left. \left\{ \epsilon_{SAk}, \epsilon_{SRk}, \epsilon_{Jk}, \epsilon_{JRk} \right\}_{k\in\mathcal{K}}  \right\}.
	\end{aligned}
	\end{equation}
	Note rigorously, the elements of the state are associated with time instant $ t $ as the argument, which is omitted for notation simplicity. Then, all possible states constitute the state space as $ \bm{s}\left(t\right) \in \mathcal{S} $, $ \forall t\:\in\mathcal{T} $.
	\item Action space: An action is defined as the deployment update of aerial platform while learning, i.e., $ \bm{a}\left(t\right) = \left[w_A^{\text{(x)}}\left(t\right) ,w_A^{\text{(y)}}\left(t\right) \right] - \left[w_A^{\text{(x)}}\left(t-1\right) ,w_A^{\text{(y)}}\left(t-1\right) \right] $ at time instance $ t\in\mathcal{T} $.
	\item Reward function: The reward function is defined as the change of robust secrecy rate as compared with that in previous time, on condition of current state and action, denoted by $ r\left(t\right) = R_S\left(t\right) - R_S\left(t-1\right) $. This reward function is defined in consistence with the definition of action and the transmission strategy is obtained through the Alg.~\ref{alg:1} to assist the evaluation of secrecy.
\end{itemize}

Given the MDP model above, the aerial platform as the agent in MDP learns to find the desired deployment. As for learning, the agent determines the action in current space according to the policy given as $ \bm{\mu}: \mathcal{S} \mapsto \mathcal{A} $. Then, the system state evolves to a new state as $ \mathcal{S} \times \mathcal{A} \mapsto \mathcal{S} $. Meanwhile, the agent obtains an instantaneous reward as $ R_S: \mathcal{S} \times \mathcal{A} \mapsto \mathbb{R}_+ $. Through the learning process, the agent intends to maximize the long-term expected reward defined as $ \Gamma = \sum\nolimits_{t=0}^{T} \nu^t r\left(t\right)$, where $ \nu\in(0,1) $ is the discount factor.

\subsection{DDPG-Based Algorithm}

As the aerial deployment issue is investigated within a continuous area, we adopt the deep deterministic policy gradient (DDPG) approach that tackles problems in continuous action space~\cite{Xiong2}. The DDPG framework has an actor-critic network structure, where the actor network observes the current state and produces an action based on the strategy and the critic network provides an evaluation regarding the action. Besides the operations in the evaluation networks noted before, the DDPG framework also incorporates the target networks integrating the experience replay. The replay buffer helps reduce the correlation of data samples and the delayed strategy updates in the target network improve the stability of the algorithm implementation.

\begin{algorithm}[t] \label{alg:2} \scriptsize
  \caption{DDPG for aerial deployment}
  Initialize the actor and critic networks with parameters $ \bm{\omega}^Q $ and $ \bm{\omega}^{\bm{\mu}} $, and copy them to the target networks as $ \bm{\omega}^{Q^{\prime}} $ and $ \bm{\omega}^{\bm{\mu}^{\prime}} $\;
  Initialize the replay buffer as $ \mathcal{D} $\;
  \For{ episode: 1 to $ \mathtt{maxEpisode} $ } 
  {
  	Initialize a random noise set for action exploration, denoted as $ \mathcal{O} $\;
  	Initialize the environment with state $ \bm{s}_0 $\;
  	\For{epoch-$t$: 1 to $ \mathtt{maxEpoch} $}
  	{
  		The agent selects an action as $ \bm{a}\left(t\right) = \bm{\mu}\left(\bm{s}\left(t\right) | \bm{\omega}^{\bm{\mu}} \right) + o\left(t\right) $, where $ o\left(t\right) \in\mathcal{O} $ is the noise for exploration\;
  		Take the selected action, obtain current reward as $ r\left(t\right) $, and update the system state as $ \bm{s}^{\prime}\left(t\right) $\;
  		Store the transition $ \left( \bm{s}\left(t\right), \bm{a}\left(t\right), r\left(t\right), \bm{s}^{\prime}\left(t\right) \right) $ in $ \mathcal{D} $\;
  		System state updates\;
  		\If{Sufficient transitions collected in $ \mathcal{D} $}
  		{
  			Randomly constitute a mini-batch of $ D $ transitions from $ \mathcal{D} $ as $ \left[\left( \bm{s}^d\left(t\right), \bm{a}^d\left(t\right), r^d\left(t\right), \bm{s}^{\prime d}\left(t\right) \right) \right]_{d=1,\cdots,D} $\;
  			Determine \textit{Q}-value according to~(\ref{eq:Q})\;
  			Update $ \bm{\omega}^Q $ by minimizing the loss function in~(\ref{eq:loss})\;
  			Update $ \bm{\omega}^{Q^{\prime}} $ according to mini-batched policy gradient as~(\ref{eq:grad})\;
  			Update the target networks with soft update method\;
  		}
  	}
  }
\end{algorithm}

The DDPG-based deployment algorithm is specified in Alg.~\ref{alg:2}, where the main operations are elaborated as follows. We first construct the evaluation actor and critic networks with parameters $ \bm{\omega}^Q $ and $ \bm{\omega}^{\bm{\mu}} $, respectively, which is then copied as the initial target networks. Also, the environment is specified based on the communication system state. Then, at the training stage, at each epoch-$ t $ with state $ \bm{s}\left(t\right) $, the agent selects and takes an action $ \bm{a}\left(t\right) $, based on current policy $ \bm{\mu}, $ along with an random noise. Meanwhile, the taken action produces an instantaneous reward as $ r\left(t\right) $ and an updated state as $ \bm{s}^{\prime}\left(t\right) $. The transition tuple $ \left( \bm{s}\left(t\right), \bm{a}\left(t\right), r\left(t\right), \bm{s}^{\prime}\left(t\right) \right) $ is stored in the replay buffer. After sufficient rounds of training to fill the replay buffer, we then randomly pick $ D $ groups of transition as a mini-batch for learning. Specifically, the evaluation critic network is trained by minimizing the loss function defined as
\begin{equation} \label{eq:loss}
	\mathsf{L}\left(\bm{\omega}^Q\right) = \frac{1}{D}\sum\limits_{d=1}^D\left( \zeta^d - Q\left( \bm{s}^d\left(t\right), \bm{a}^d\left(t\right) \right) \right)^2,
\end{equation}
where
\begin{equation} \label{eq:Q}
	\zeta^d = r^d + \nu Q^{\prime}\left( \bm{s}^{\prime d}\left(t\right), \bm{a}^{\prime d}\left(t\right) \right)  \left|_{\bm{a}^{\prime d}\left(t\right) = \bm{\mu}\left( \bm{s}^{\prime d}\left(t\right) \right)} \right.,
\end{equation}
with $ Q $ and $ Q^{\prime} $ being the action-value function. Meanwhile, the evaluation actor network parameters are updated according to gradients as
\begin{equation} \label{eq:grad}
\begin{aligned}
        \Delta_{\bm{\omega}^{\mu}} J = \frac{1}{D}\sum\limits_{d=1}^D & \nabla_{\bm{\omega}^{\mu}} \bm{\mu}\left( \bm{s}^{d}\left(t\right) \right) \cdot\\
         &\nabla_{\bm{a}^{ d}\left(t\right)} Q \left( \bm{s}^{ d}\left(t\right), \bm{a}^{ d}\left(t\right) \right) \left|_{\bm{a}^{ d}\left(t\right) = \bm{\mu}\left( \bm{s}^{ d}\left(t\right) \right)}  \right..
\end{aligned}
\end{equation}
Then, the target network parameters are updated according to the soft update rule to improve the stability of the learning process.

When the neural networks are well-trained, the agent selects the action based on the network output at each stage, given certain initial communication network status. Then, the final convergence provides a learnt deployment for the aerial platform to assist the robust secure transmissions.

\subsection{Implementation Issue}

As the proposed DDPG-learning algorithm (Alg.~\ref{alg:2}) incorporates the BCD-based transmission algorithm (Alg.~\ref{alg:1}) providing intermediate results, we discuss the overall algorithm implementation here. First, regarding the initialization, the aerial platform can be placed at any random spot within the region, and the artificial noise covariance and reflection coefficients can also be randomly initialized as long as the constraints in~(\ref{eq:innerProb}) on jamming power and modulus are satisfied. Then, the channel information is required to conduct the algorithm. In this regard, we can use the properly designed training sequences to efficiently estimate the cascaded channels between the legitimate pair~\cite{Xing}. For passive eavesdroppers, we can simply use their location information to estimate the channel due to the line-of-sight-dominated air-ground transmissions in our considered scenarios. Also, the uncertainty-associated error bounds can be obtained based on historical data. Afterwards, the proposed algorithms are conducted based on the collected information. Technically, the computation can be carried out at the aerial platform which usually has the processors and energy source. Then, the aerial deployment, ARIS reflection, and jamming can be readily applied at the aerial platform, while the results are also feed back to the fixed RIS to update the reflection coefficients.

\section{Simulation Results} \label{sec6}

In this section, we present simulation results to show the performance of the proposed robust secure transmission scheme. Specifically, we consider a ground area of 400$\times$400 (distance in meters and the same afterward), where the transmitter is located at the origin and the legitimate receiver is located at (350, 0). There are three eavesdroppers randomly located in a circle area centered at (300, 300) with a radius of 50, noted as the eavesdropping area. Also, the fixed RIS is located at (100, 150) with a height of 50, while the height of the ARIS is assumed of 150. As the signal propagation in the system is concerned either with the fixed RIS or the aerial platform, we adopt the Rician channel model. In particular, the channels associated with the fixed RIS are assumed of a Rician factor of 3~dB, while the channels related to the aerial platform are of a Rician factor of 10~dB, due to the fact that the aerial platform locates higher than the fixed RIS. Meanwhile, the path loss exponents for the air-ground channels concerning the fixed RIS and ARIS are 2.6 and 2.2, respectively. The air-to-air links between the aerial jammer and the fixed RIS have a path loss exponent of 2. The path loss at the reference distance is 20~dB for all channels. The transmit power at the source node is 30~dBm, and the aerial jammer has 4 antennas with a maximum jamming power of 25~dBm. The background noise power is -110~dBm. The fixed RIS and ARIS both have 50 reflecting elements. Regarding the channel uncertainties, we define the uncertainty coefficient denoted by $ \delta $, given as $ \epsilon_{X} = \delta \left\| \hat{\bm{h}}_{X} \right\| $, where $ X \in \left\{ SAk, SRk, Jk, JRk \right\} $ with $ k\in\mathcal{K} $. The uncertainty coefficient is assumed of 0.01. Moreover, the parameters for reinforcement learning are detailed as follows. The number of episodes is 2,000. The replay buffer size and batch size are 20,000 and 256, respectively. The learning rates for the actor network and critic network are 0.0001, and the soft update coefficient is 0.005. The discount factor is 0.95.

\begin{figure}[t] 
  \centering
  \includegraphics[width=9.0cm]{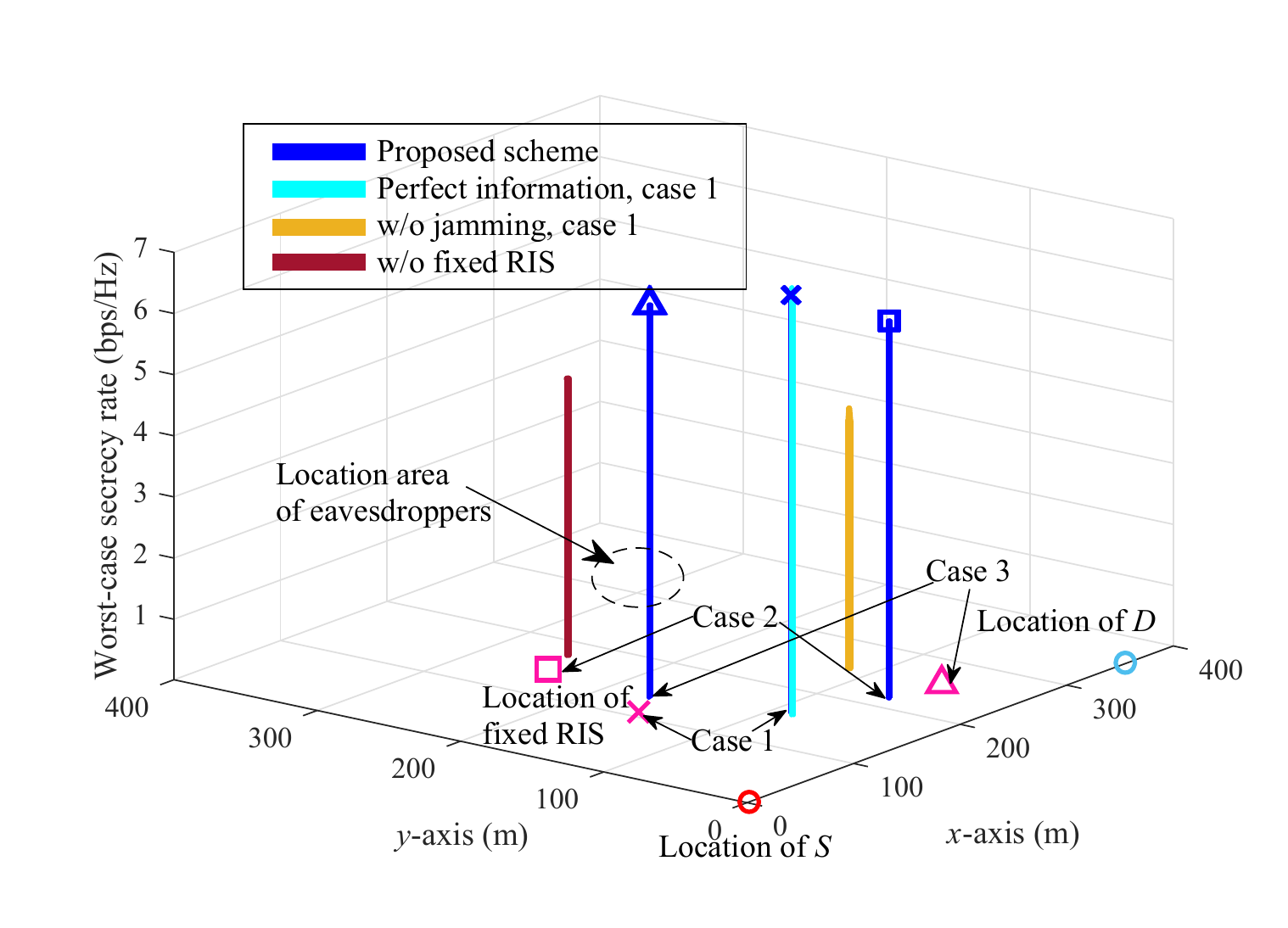}
  \caption{Illustrations of ARIS deployment and secrecy performance.}
  \label{fig:deploy}
\end{figure}

Fig.~\ref{fig:deploy} demonstrates the ARIS deployment along with the achieved robust secrecy rate. In this figure, the bar location indicates ARIS deployment and the bar height corresponds to the achieved robust secrecy rate. Three cases are illustrated with the fixed RIS deployed at (100, 150), (150, 250), and (250, 50) under Case 1,2,3, which lead to the aerial deployment at (161, 89), (218, 63), and (136, 169), respectively. As we can see, the deployment results imply that usually one RIS locates closer to the eavesdroppers while the other closer to the legitimate receiver. It can be explained that one RIS nearer to the legitimate receiver helps enhance the reception, while the other nearer to the eavesdroppers strengthens the active jamming, by either direct jamming (through aerial jamming) or reflective jamming (through the fixed RIS). Also, the induced robust secrecy rates under different cases are rather close, indicating that the flexible deployment of the ARIS can effectively compensate for the performance under different fixed-RIS deployments. Moreover, under Case-1 deployment, we show the results under perfect CSI, without jamming or fixed RIS. As we can see, when without jamming, the ARIS deployment locates farther to the fixed RIS as compared with the case with jamming, since the aerial platform no longer needs the fixed RIS to enhance the active jamming. When there is no fixed RIS, the aerial platform locates nearer to the eavesdroppers as active jamming can be more direct and effective in defending against eavesdropping. Further, for Case-1, we find the optimal deployment through global search instead of reinforcement learning, inducing the aerial deployment at (157, 85) with an achieved robust secrecy rate at 6.93 bps/Hz. In contrast, the aerial deployment through proposed learning is at (161, 89) with a robust secrecy rate of 6.83 bps/Hz. Note that the results under global search are not explicitly shown in the figure as they are rather close to the existing demonstration. The results indicate that our proposal can effectively solve the problem while approaching the optimum.

\begin{figure*}[t] 
  \centerline{
  \subfigure[Performance under different uncertainties.]{
    \label{fig:robustEveLoc} 
    \includegraphics[width=7.5cm]{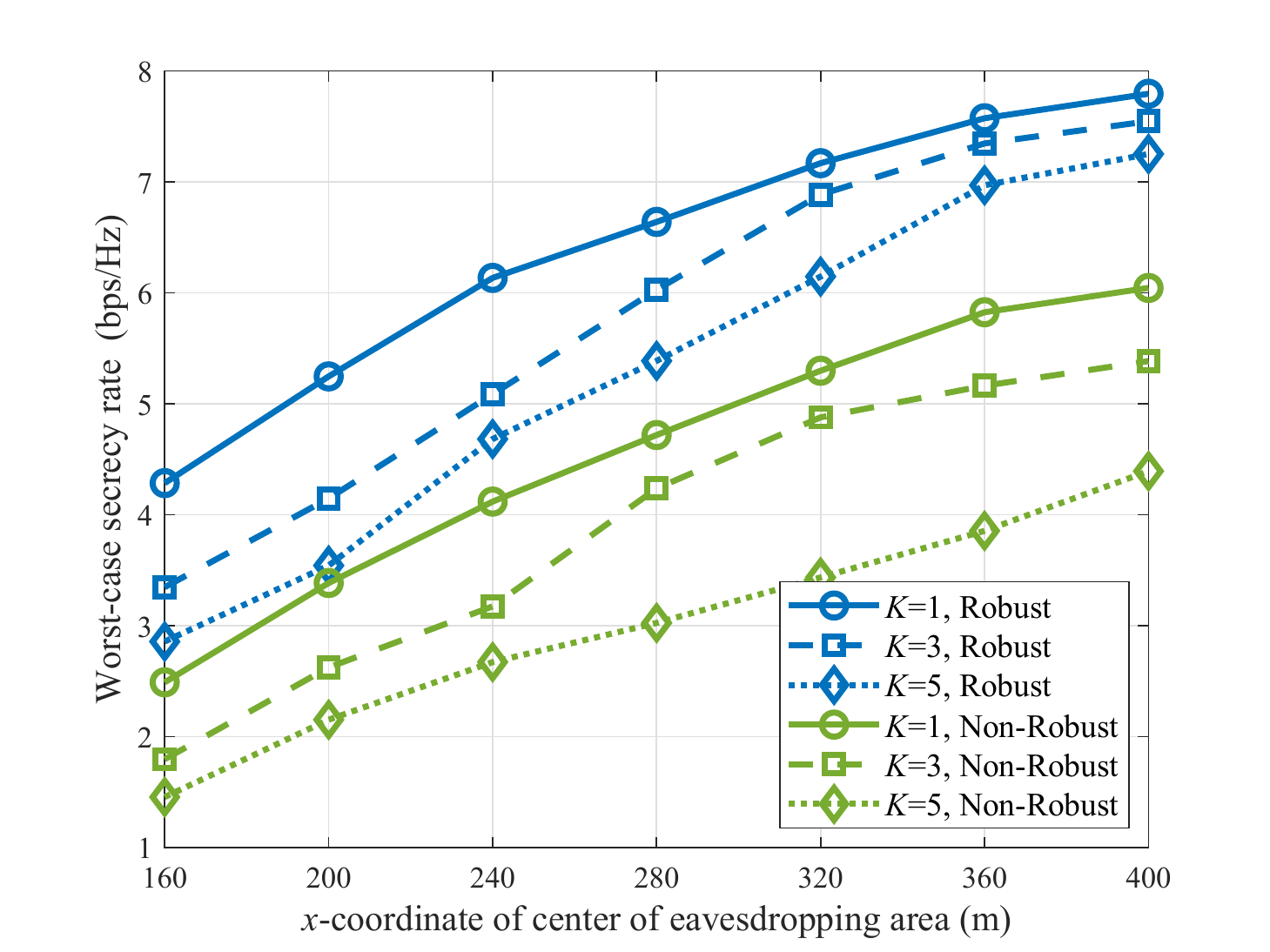}}
  \hspace{42pt} 
  \subfigure[Performance comparison under different schemes.]{
    \label{fig:PerfEveLoc} 
    \includegraphics[width=7.5cm]{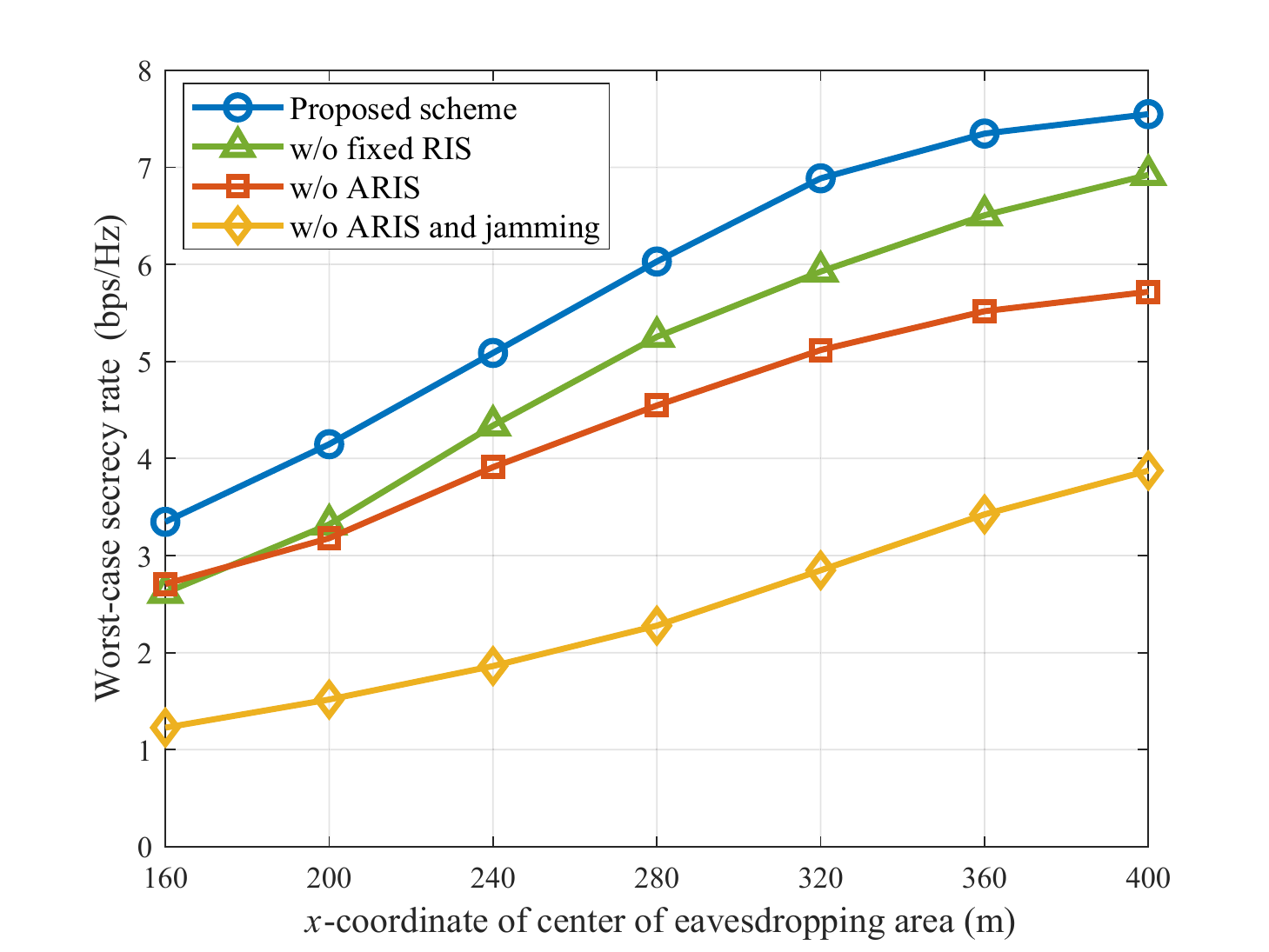}}
  }
  \caption{Performance with respect to the area center location of the eavesdroppers.}
  \label{fig:EveLoc} 
\end{figure*}

\begin{figure*}[t] 
  \centerline{
  \subfigure[Performance under different uncertainties.]{
    \label{fig:robustPwr} 
    \includegraphics[width=7.5cm]{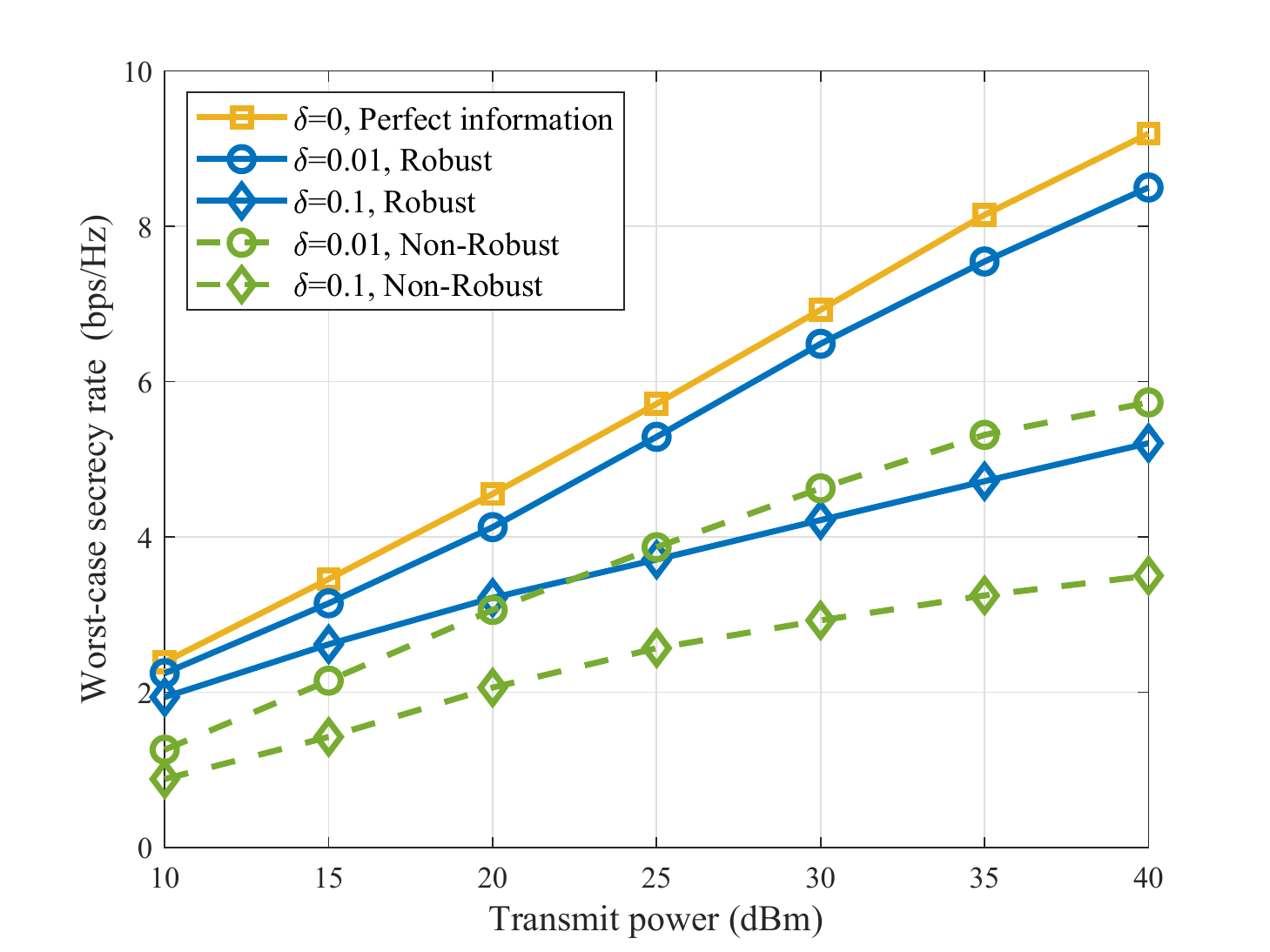}}
  \hspace{42pt} 
  \subfigure[Performance comparison under different schemes.]{
    \label{fig:perfPwr} 
    \includegraphics[width=7.5cm]{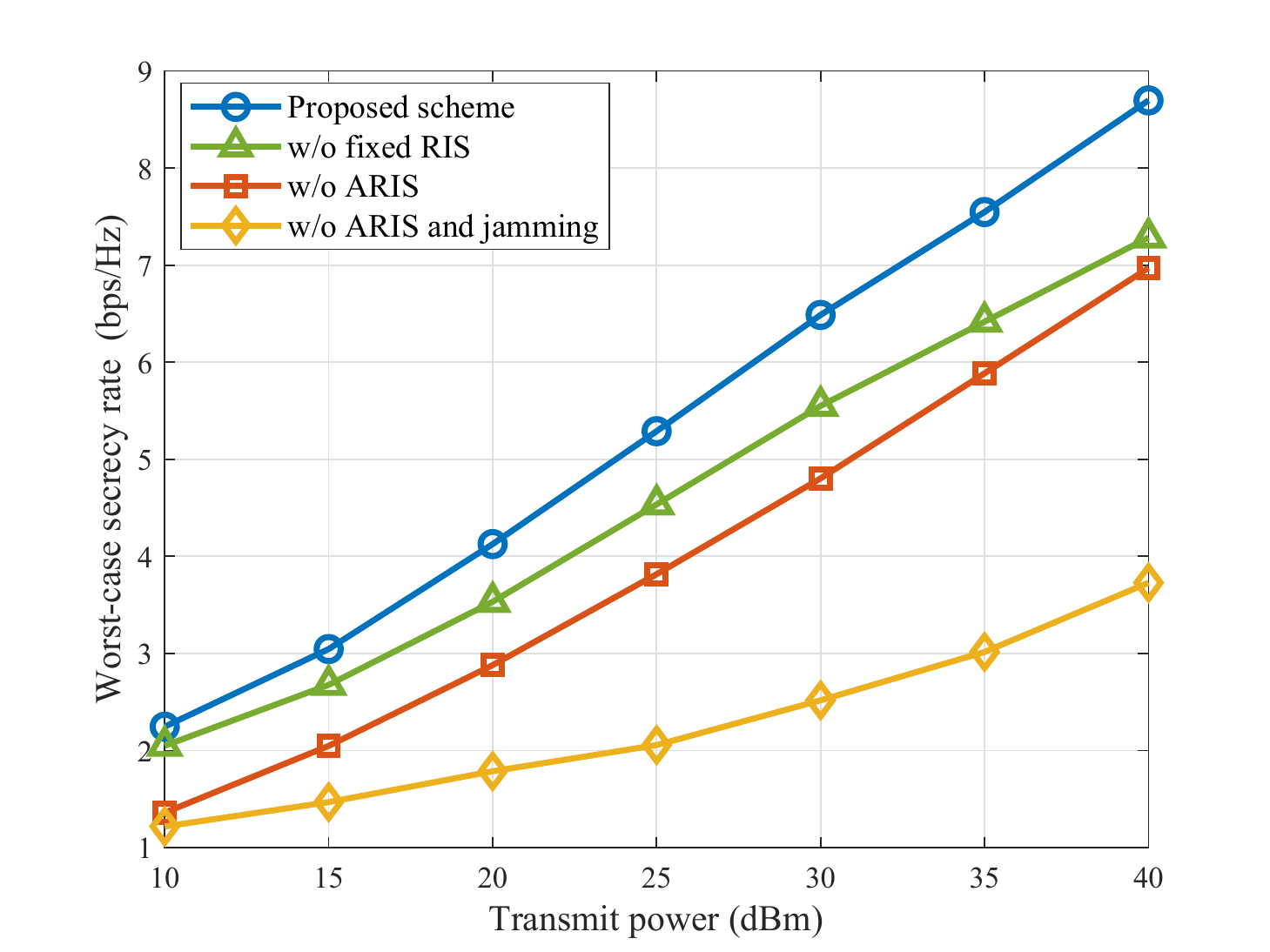}}
  }
  \caption{Performance with respect to the transmit power.}
  \label{fig:pwr} 
\end{figure*}

Fig.~\ref{fig:EveLoc} shows the performance with different locations of the eavesdropping areas. In Fig.~\ref{fig:robustEveLoc} showing the robustness performance, we can see that the worst-case secrecy rate is higher with a smaller number or farther location of the eavesdroppers. Also as expected, the robust scheme has better performance as compared with the non-robust one under the worst case. Moreover, when the eavesdroppers locate farther, the performance superiority of the robust scheme as compared with the non-robust one becomes more evident. In Fig.~\ref{fig:PerfEveLoc}, we show the performance comparison under different schemes. Besides the similar trend as that in Fig.~\ref{fig:robustEveLoc}, we can see our proposal outperforms the baselines. Moreover, from the cases without fixed RIS or ARIS, we can see that the ARIS is more effective in defending against eavesdropping attacks, due to its flexible deployment. Further, when there is neither ARIS nor jamming, i.e., removing the aerial platform, the performance with one single RIS under fixed deployment can be significantly undermined.

Fig.~\ref{fig:pwr} shows the performance concerning the transmit power. In Fig.~\ref{fig:robustPwr} showing the robustness, the secrecy rate increases with higher transmit power and smaller uncertainties. Particularly, we can see that when the uncertainty is larger, the transmission behavior becomes more conservative to tackle the worst case, and thus the speed of secrecy rate increase is slower as compared with that with smaller uncertainties. Accordingly, the performance gap between the robust scheme and the non-robust one is larger when the uncertainty is smaller. Moreover, compared with the case with perfect information, the performance can be evidently degraded when considering the channel uncertainties. Fig.~\ref{fig:perfPwr} compares the performance under different proposals, showing similar trends as those in Fig.~\ref{fig:PerfEveLoc}. We can also observe that when the transmit power is higher, the advantage of our proposal becomes more significant as compared with the baselines.

\begin{figure*}[t] 
  \centerline{
  \subfigure[Performance under different uncertainties.]{
    \label{fig:robustNumEle} 
    \includegraphics[width=7.5cm]{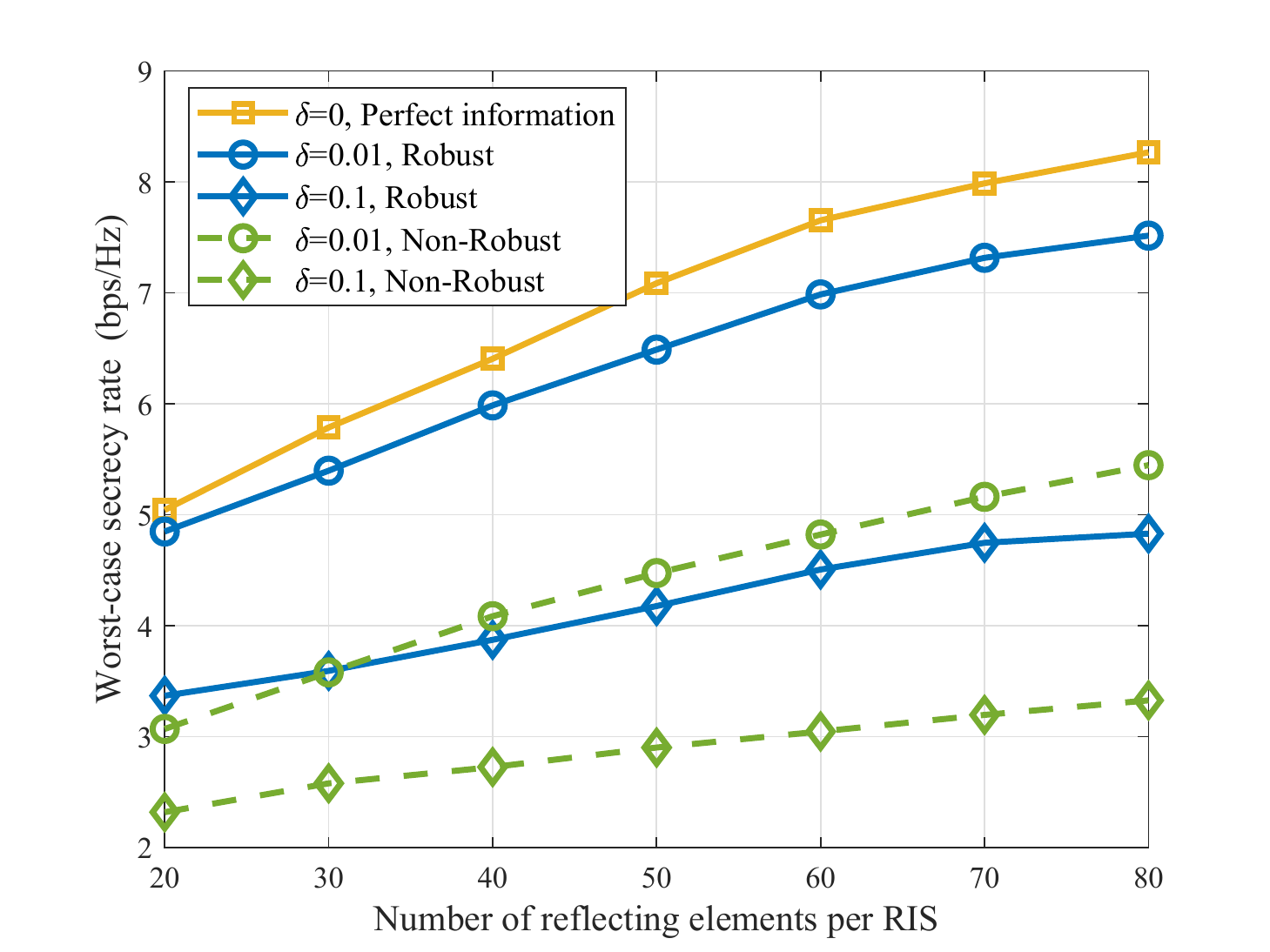}}
  \hspace{42pt}  
  \subfigure[Performance comparison under different schemes.]{
    \label{fig:perfNumEle} 
    \includegraphics[width=7.5cm]{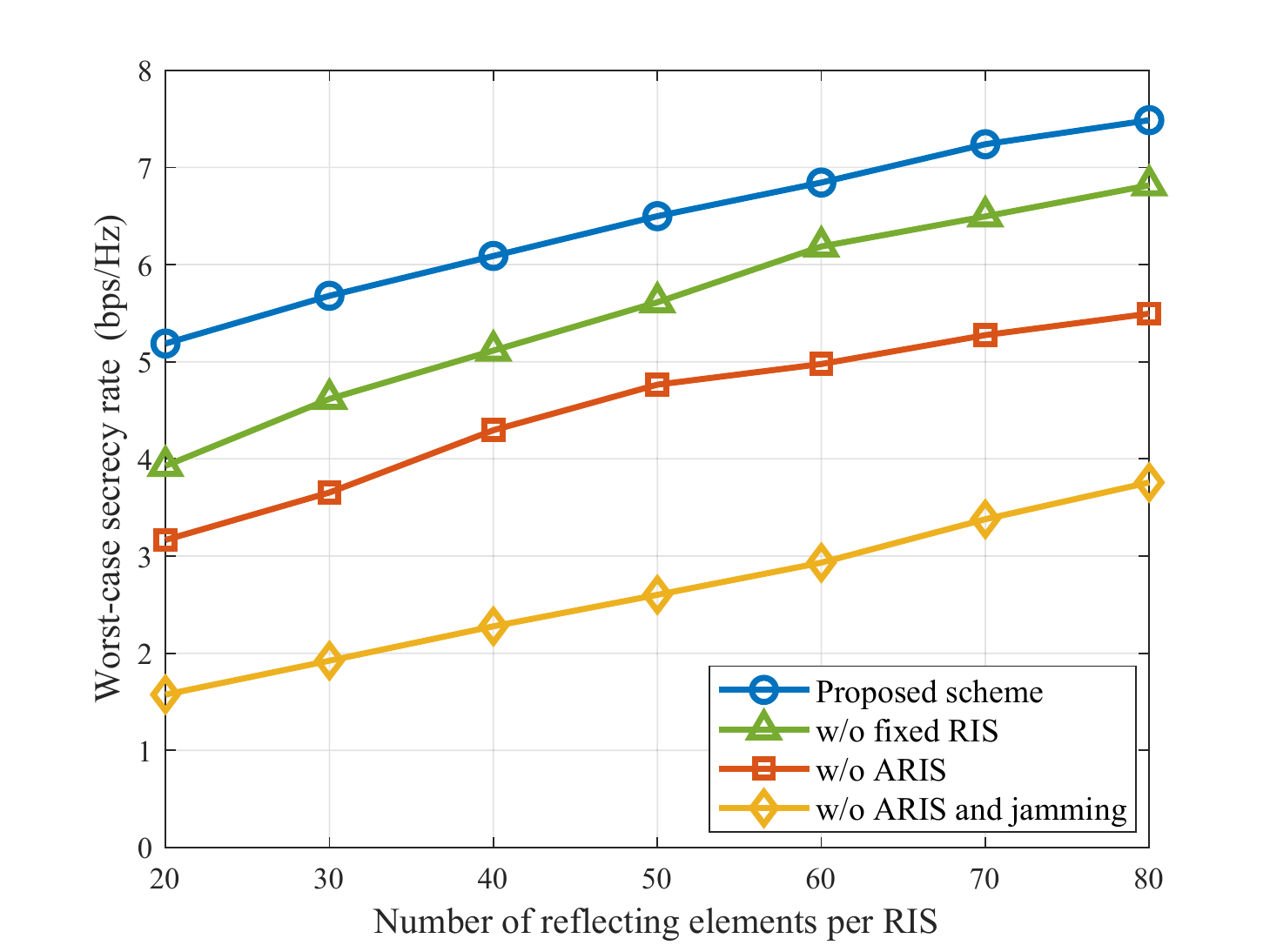}}
  }
  \caption{Performance with respect to the number of reflecting elements at each RIS.}
  \label{fig:numEle} 
\end{figure*}

Fig.~\ref{fig:numEle} shows the performance considering the number of reflecting elements at the RISs. Fig.~\ref{fig:robustNumEle} for robustness and Fig.~\ref{fig:perfPwr} are of similar trends as those in Figs.~\ref{fig:robustPwr} and~\ref{fig:perfPwr}, respectively. While for Figs.~\ref{fig:pwr} and~\ref{fig:numEle}, we can see that the worst-case secrecy rate increases almost linearly with the exponential increase of transmit power, while with the linear increase of the number of reflecting elements at the RISs. In this regard, we can see that the application of RISs in wireless networks can effectively compensate for the security performance if the transmit power is bottlenecked in wireless networks. Moreover, in Fig.~\ref{fig:pwr}, the differences in achieved robust secrecy rates under different proposals are enlarged when the transmit power increases, while in Fig.~\ref{fig:numEle}, the differences among different proposals almost remain constant. This indicates that the reflection can magnify the effect of security enhancement through increased power at the source.

\begin{figure}[t]
  \centering
  \includegraphics[width=7.5cm]{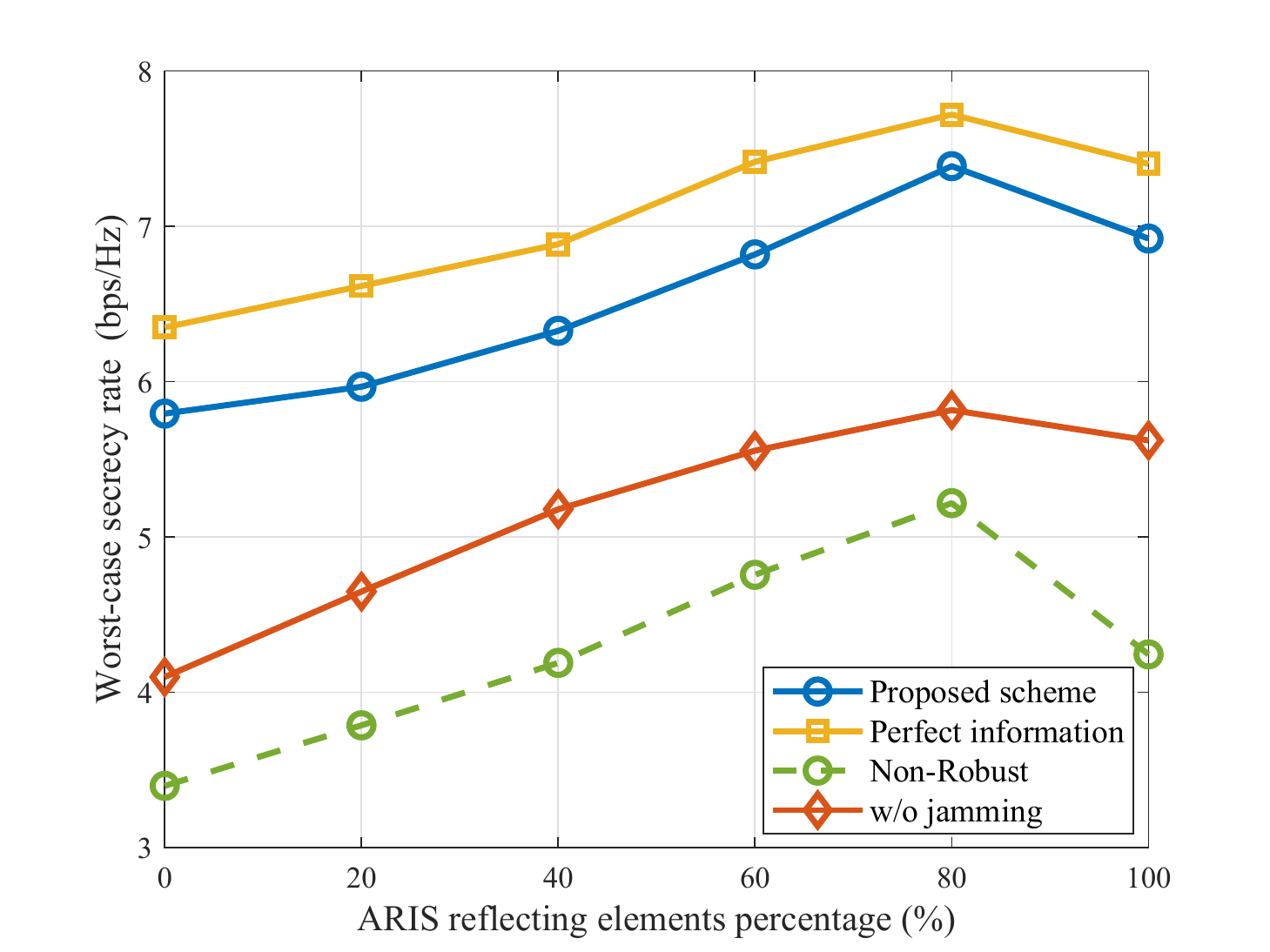}
  \caption{Performance comparison under different settings of the fixed RIS and ARIS.}
  \label{fig:perct}
\end{figure}

In Fig.~\ref{fig:perct}, we consider that the total number of reflecting elements at the RISs is fixed at 100, while evaluating the performance with different settings of elements at the fixed RIS and ARIS. Similarly, we can see that our proposal outperforms the non-robust schemes or the case without jamming, yet is inferior to the case without uncertainties. More importantly, we can see that there exists a tradeoff in distributing the reflection capability in two RISs, and the collaboration between two RISs generally improves the security as compared with the cases to use one single. Also, from the leftmost and rightmost cases in the figure, corresponding to solely using the fixed RIS and ARIS, we can see that the cases with ARIS have superior performance to the other, indicating that the flexible deployment brings ARIS evident advantages. Moreover, we can see that the configuration achieving the highest robust secrecy rate is that the ARIS is equipped with 80\% of the total reflecting elements, indicating that the RIS with optimized deployment deserves higher reflecting capability to achieve the best performance.

\begin{figure}[t]
  \centering
  \includegraphics[width=7.5cm]{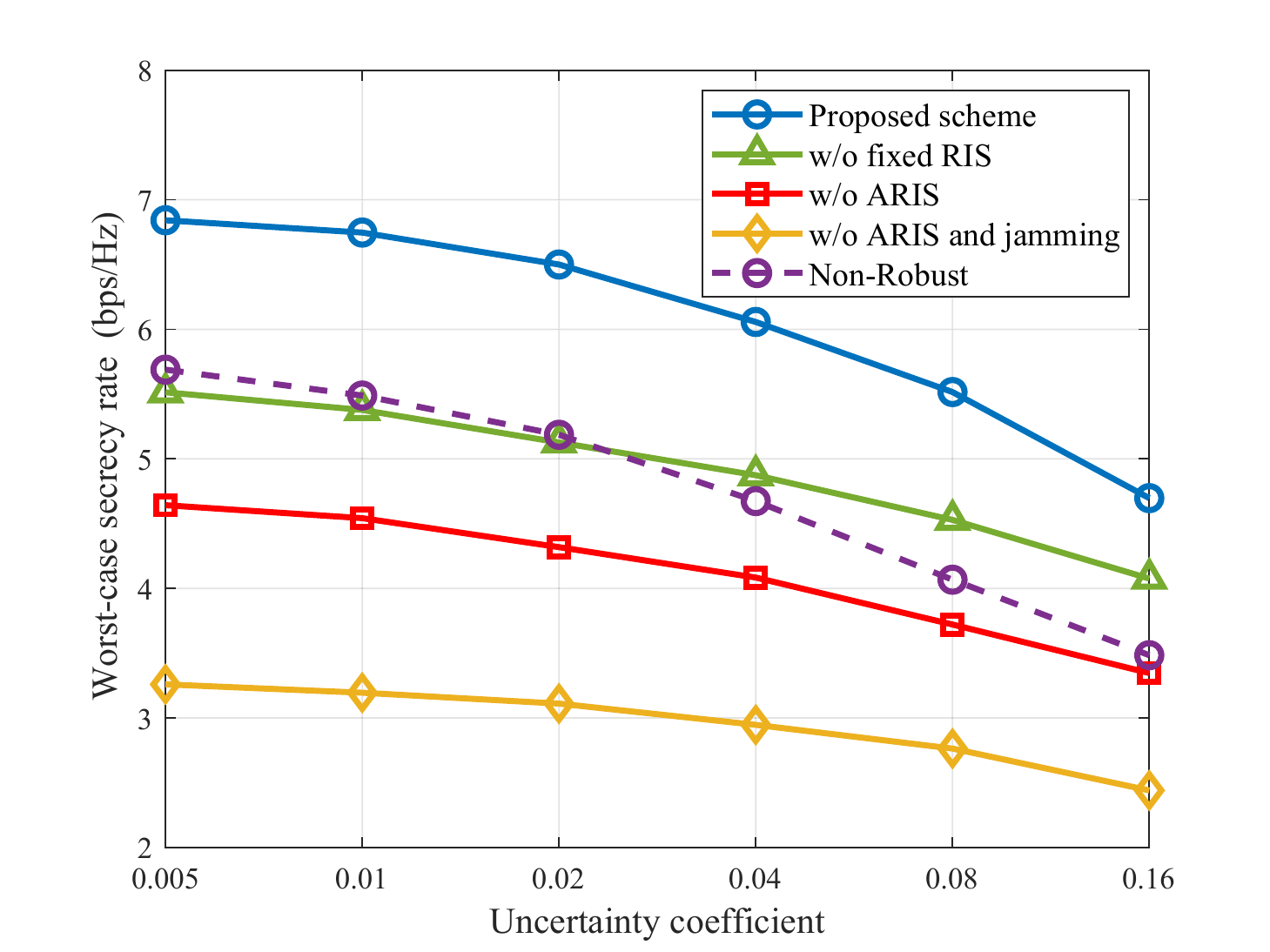}
  \caption{Performance comparison under different uncertainty coefficients.}
  \label{fig:unc}
\end{figure}

In Fig.~\ref{fig:unc}, we show the performance considering different uncertainty levels. Generally, the security performance is downgraded with higher uncertainties for all schemes, while our proposal outperforms the baselines. Also, the robust secrecy rate decreases faster under our proposal compared with the cases without one single RIS or jamming. This is because, when without RIS or jamming, the corresponding channels for reflection or jamming related to the eavesdroppers no longer exist, and so are the associated uncertainties. In this regard, the overall channel information uncertainty in the system is reduced and thus the security performance loss due to uncertainties is alleviated.

Overall, the numerical results have shown the effectiveness of our proposed scheme for security provisioning while tackling uncertainties. Particularly, we can see that the ARIS with flexible deployment generally outperforms the conventional fixed RIS deployment, where the security performance can be further enhanced with cooperative aerial jamming. Moreover, the joint use of fixed RIS and ARIS can dynamically adapt to different network topologies, as the flexible deployment of ARIS can effectively compensate for the potentially unfavorable location of the fixed RIS.

\section{Conclusion} \label{sec7}

In this paper, we propose to exploit aerial reflection and jamming to enhance wireless security, where robust security is proposed to address the channel uncertainties. Specifically, we employ robust optimization approaches to tackle the reflection and jamming, and aerial deployment is obtained through deep reinforcement learning. Results show that the proposed scheme can effectively combat the uncertainties to achieve robust security under the worst case. Also, the ARIS with flexible deployment is more effective compared with fixed RIS in terms of security provisioning, and the collaborative operation between the fixed RIS and ARIS can significantly improve the security performance.

\bibliographystyle{IEEEtran}
\bibliography{main}

\end{document}